\documentclass[12pt]{article}
\usepackage[utf8]{inputenc}
\usepackage[font={small,it}]{caption}
\usepackage{geometry}
\usepackage{graphicx} 
\graphicspath{{figures/}}
\usepackage{enumerate}
\usepackage{graphicx}
\usepackage{amsmath}
\usepackage{placeins}
\usepackage{bbm,bm}
\usepackage[english]{babel}

\usepackage{caption}
\usepackage{algorithm}
\usepackage{algpseudocode}
\usepackage{amssymb}
\usepackage{adjustbox}
\usepackage{rotating}
\usepackage[colorinlistoftodos]{todonotes}
\usepackage{ulem}
\usepackage[square,sort,comma,numbers]{natbib}

\begin{document}

\title{Change-point detection using spectral PCA for multivariate time series}
\author{Shuhao Jiao\footnote{Statistics Program, King Abdullah University of Science an Technology, Saudi Arabia; \texttt{shuhao.jiao@kaust.edu.sa}} \phantom{.},
	Tong Shen\footnote{Department of Statistics, UC Irvine, USA; \texttt{tong@uci.edu}} \phantom{.}, 
	Zhaoxia Yu\footnote{Department of Statistics, UC Irvine, USA; \texttt{zhaoxia@ics.uci.edu}} \phantom{.},
	Hernando Ombao\footnote{Statistics Program, King Abdullah University of Science an Technology, Saudi Arabia;  \texttt{hernando.ombao@kaust.edu.sa}}
}
\maketitle



	\begin{abstract}
		We propose a two-stage approach Spec PC-CP to identify change points in multivariate time series.
In the first stage, we obtain a low-dimensional summary of the high-dimensional time series by Spectral Principal Component Analysis (Spec-PCA).  In the second stage, we apply cumulative sum-type test on the Spectral PCA component using a binary segmentation algorithm. Compared with existing approaches, the proposed method is able to capture the lead-lag relationship in time series.
Our simulations demonstrate that the Spec PC-CP method performs significantly better than competing methods for detecting change points in high-dimensional time series. The results on epileptic seizure EEG data and stock data also indicate that our new method can efficiently {detect} change points corresponding to the onset of the underlying events. 
	\end{abstract}

\section{Introduction}
Detecting change point in time series is an important goal in many scientific disciplines such as medicine, financial markets, and climate (Scheffer et al., 2009). For instance, much effort has been dedicated to capture the sudden changes in brain due to epilepsy. The analysis of epilepsy helps people understand disruptions in normal brain functioning and develop more precise diagnosis, improve therapy, and develop effective early-warning systems for onset of seizure activity (see, for example, 
Schr\"oder and Ombao 2019; Euan et al., 2019). 
Epileptic seizure prediction has been a notably challenging problem (Mormann et al., 2006).  Recent studies showed that prediction might be feasible by analyzing ectroencephalography (EEG) time series ( Chisci et al., 2010). These automatic algorithms for seizure detection aim to provide real-time detection and onset warnings, which may greatly improve clinical and monitoring diagnosis. 


In the financial field, Aue and collaborators (Aue, Horv\'ath and Reimherr,  2009; 
Aue, H\"ormann and Horv\'ath, 2009) developed rigorous change-point detection methods. They introduced an asymptotic test procedure to assess the stability of volatilities and cross-volatilites.
It is of prime importance to detect the potential change in stock market and during financial crisis, which enables people to better understand the relationship between financial events and stocks. Choudhry (1996) studied the 1987 financial crisis. Generalized autoregressive conditional heteroskedasticity-in-mean (GARCH-M) model was used to investigate the volatility and changes in volatility for emerging markets before and after the crisis. 
In Dooley and Hutchison (2009), the subprime crisis circa 2008 was analyzed. A long decline was observed in U.S. equities at the start of subprime crisis in mid-2007 through September 2008. There was also a dramatic change in U.S. stock market in September 2008 with increased volatility after the crisis.

In the literature, both parametric and nonparametric methods have been proposed to detect change points using either univariate or multivariate time series (Chen at al., 2010; Davis et al., 2006, Chen and Gupta, 2012). Parametric models are efficient if the model assumptions are correct. Kirch et al. (2015) developed an approach where a score type of test statistic based on vector autoregressive (VAR) models was used for identifying change points. Nonparametric change points are also used for detecting changes in the spectral characteristics of the time series such as those based on cumulative sum-type tests (Adak, 1998; Ombao et al., 2005;  Terrien, Germain, Marque and Karlsson, 2013). For example, Ombao et al. (2005) developed a procedure for segmentation of a non-stationary time series via model selection using the SLEX library which is a collection of orthonormal time-localized Fourier waveforms. 
One limitation of these methods is they lack the interpretability for frequency as we expect to detect change points for EEG seizure on frequency bands as well. Frequency-specific change point detection can help people gain more insights in brain activity at different frequencies (Alarcon, Binnie, Elwes and Polkey, 1995; Blondin and Greer, 2011;  Schmitt, Pargeon, Frechette, Hirsch, Dalmau, and Friedman, 2012). Moreover, when we extend methods from univariate to multivariate time series case, existing methods can be computationally inefficient. 

In practice, multivariate time series are often correlated. Analyzing or modeling them jointly, rather than individually or pairwisely, allows us to take the full multivariate information into consideration  (Ombao et al., 2005); as a result, multivariate approaches are often more efficient. Besides, in many time series data, spectral analysis is often able to extract critical spectral characteristics of the data by examining different frequency bands \citep{kaplan2001macrostructural}. Motivated by these previous work, we develop 
a two-stage detection method for multivariate time series which we call the Spectral PCA change point (Spec PC-CP) method. The main advantage of
our method is that it reduces computational complexity and is
more likely to give lower reconstruction error (Wang et al., 2016) especially when there is lead-lag relationship between components of the time series.

The rest of the paper is organized as follows. In section 2 we propose the two-stage Spec PC-CP method and describe the comparison methods:  Contemporaneous Mixture Method mentioned (Wang et al., 2016), Structural Break Detection (Safikhani and Shojaje, 2017) and Sparsified Binary Segmentation (Cho and Fryzlewicz, 2015). In section 3 we conduct simulation studies 
to evaluate the performance of Spec PC-CP under several simulation settings. In section 4, we analyze epileptic seizure EEG data and stock data and show that our method can efficiently detect change points corresponding to the onset of seizure in EEG data and financial issues in stock data. In section 5 we conclude our paper to discuss the potential future directions.

	\section{The Spec PC-CP method}
	In multivariate analysis, PCA is one of the mostly used techniques to extract linear combinations that provide low dimensional summaries of multivariate data. The first principal component is given by the linear combination with the largest variance among all possible linear combinations. Each following component has the largest variance under the constraint that it is orthogonal to the preceding components. By extracting a few leading components that account for a desired proportion of the total variance, PCA is often used for visualization and dimension reduction. 
	
	\textbf{Contemporaneous Mixture Method}. Consider a zero-mean $p$-dimensional process $X(t)$. In classical PCA, the PCs as mentioned in Wang et al., (2016) can be represented as an instantaneous linear mixture of the original $p$-dimensional time series $X(t)$. The calculating procedure is as follows:

\begin{itemize}
	\item  For $X(t)$,  compute the eigenvalues-eigenvector pairs of $\Sigma^x$ as $\{(\lambda_\ell, \mathbf{e}_\ell)\}^q_{\ell=1}$, where $\lambda_1 > \lambda_2 >,...,> \lambda_q$ and $\lVert \mathbf{e}_\ell \rVert =1$. When $\Sigma^x$ is not known, we estimate ${\Sigma}^x$ by $\hat{\Sigma}^x = \frac{1}{T}\sum_{t=1}^T X(t){X(t)}^T$, where $T$ is the length of the series
	
	\item  The $\ell$-th component can be calculated by
	\begin{center}
		$u_\ell(t) = e_\ell^TX(t)$ 
	\end{center}
\end{itemize}

Although widely used in dimension reduction, classical PCA may not be optimal when there is lead-lag relationships between multiple time series. Spectral PCA, which was originally introduced in Brillinger (1964), aims to find components that explain the variance in the spectral density matrix at all frequencies rather than the contemporaneous covariance matrix. Because of the 1-1 relationship between the sequence of auto-covariance matrix at all lags and the spectral density matrix at all frequencies,
Spectral PCA is able to capture lead-lag dependence at various lags between the different time series. We will show an example in the simulation section where Spectral PCA gives a better summary of the latent source (Figure 3). An additional advantage of the Spectral PCA approach is that it allows us to examine the roles of different frequency bands. Because Spectral PCA has the advantages of both spectral analysis and PCA, it has been used to analyze multivariate time series data, (See, for example, Wang et al., 2016, Ombao et al., 2005, Ombao and Ho, 2006, B\"ohm et al., 2010 and Stoffer et al., 2002). In spectral PCA, we summarize a $p$-dimensional time series $X(t) = (X_1(t), \ldots, X_p(t))'$, by $q(\le p)$ univariate time series, each of which is a linear combination of all past, present and future observed time series, i.e, a linear filtered version of $X(t)$:

\begin{center}
	$u(t)=\sum\limits_{h=-\infty}^\infty  A(h) X(t-h)$
\end{center}
Based on $u(t)$, one can construct $X(t)$ by $X(t)=\sum\limits_{h=-\infty}^\infty B(h)u(t-h)$. Here $A(h)$ and $B(h)$ are absolutely filters with dimensions $q\times p$ and $p\times q$ respectively. 
The filters are chosen to minimize the eigenvalues of the approximation error matrix $E[(X(t)-\hat{X}(t))^\ast (X(t)-\hat{X}(t))]$. Similar to classical PCA, there are up to $p$ components. We use $u_\ell(t)$ to denote the $\ell$-th spectral PC. We apply the following algorithm to extract the first $q$ spectral PCs.


\begin{algorithm}[h]
	\caption{Spectral PCA Method for Extracting Summaries}\label{euclid}
	\begin{algorithmic}[1]
		\Function {Spectral PCA for data matrix  $X(t)$ }{}
		\For{$j=0,1,...,T-1$} 
		\State $d(\omega_j)= T^{-1/2} \sum\limits^T_{t=1} X(t) \exp(-2\pi i\omega_j t)$ (Compute Fourier coefficients(FFT)), where $\omega_j=j/T$
		\State Compute periodogram matrix $I(\omega_j)= d(\omega_j)\cdot d^\ast (\omega_j)$
		\State Estimate the spectral density matrix $\widehat{f}(\omega_j) = smooth_{r} I(\omega_{j + r})$ (use Daniel kernel with window size of length $5$)
		\State Compute the $q$ largest eigenvalues of $\widehat{f}(\omega_j)$ denoted 
		\[
		0 < \lambda_q(\omega_j) < \ldots, \lambda_1(\omega_j)
		\]
		\State Extract the eigenvectors $V_{\ell}(\omega_j)$ that correspond to the $\ell$-th largest eigenvalues, 
		$\ell = 1, \ldots, q$
		\EndFor
		\State Compute the filter $b_\ell(h)=\sum\limits_j V_\ell^\ast(\omega_j)*\exp (2\pi i\omega_jh)$ (compute the filter that correspond to th $\ell$-th largest eigenvalue) for $\ell = 1,...,q$
		\State Obtain the first $q$ spectral PCs $u_{t,\ell}=\sum\limits_{h=-R}^R b_{h,\ell}*X_{t-h}$ for $\ell = 1,...,q$\\
		\Return $u(t)=u_1(t),...,u_q(t)$
		\EndFunction
	\end{algorithmic}
\end{algorithm}
\FloatBarrier

In the algorithm of Spectral PCA, we first compute the eigenvectors of the smoothed periodogram (an estimation of spectrum) for each frequency. After that we calculate time-varying filter based on the eigenvectors. Finally by applying the filter to the original data, we are able to get the first $q$ Spectral PCs, which are low dimensional summaries.
The low dimensional summaries $u_{\ell}(t)$ can be thought as the frequency domain analog of the $\ell$-th principal component in classical PCA. In this article, we calculate the first three spectral PCs ($\ell = 1,2,3$) in the algorithm to conduct change point analysis.

\subsection{Stage II: Identifying change-points from a Spectral PC using the binary segmentation algorithm}
Spectral PCA was proposed to conduct dimension reduction for stationary time series (Brillinger 1981; Shumway and Stoffer, 2017). Time series with potential change points, such as EEG and stock data
collected over a long period time, are naturally non-stationary. Thus, it is necessary to consider segmentation methods when identifying change points. For example, the SLEX library proposed by 
Ombao et al. (2005) consists of localized waveform where localization is achieved by binary segmentation so that each time block is approximately stationary.

When combined with binary segmentation, the CUSUM statistics can consistently detect multiple change-points in a recursive manner (Cho and Fryzlewicz, 2012). 
In this paper, the CUSUM statistic is applied to the spectrum of the time blocks. For each time block,
we first compute the smoothed periodogram of the $\ell$-th spectral PC $u_{\ell}(t)$ (See algorithm 1(4,5,6))
Similar to Schr\"oder and Ombao (2019),
we let $\hat{f}(\tilde{t},\omega_j)$ denote the estimate of the spectrum of $u_{\ell}(t)$ within the $\tilde{t}$-th block at frequency $\omega_j$.
Figure 2 shows an example of a multivariate time series($T=1,000$ and $p=20$ channels) with two change points at $t=400$ and $700$. The time series is a mixture of AR(2) latent sources with different frequency bands. Between the two change points, the weight of the alpha band (8-12hz) is lower while   the weights of delta (0-4hz) and gamma (30-45hz) band remain constant across all time blocks. The estimated spectrum of the first spectral PC is calculated at each time block with length of 100. It is clear that the estimates agree with the latent source. Two change points are detected using the first spectral PC, which agrees with the truth.
\begin{figure}[h]
	\centering
	\begin{tabular}{c c}
		\includegraphics[scale=0.5]{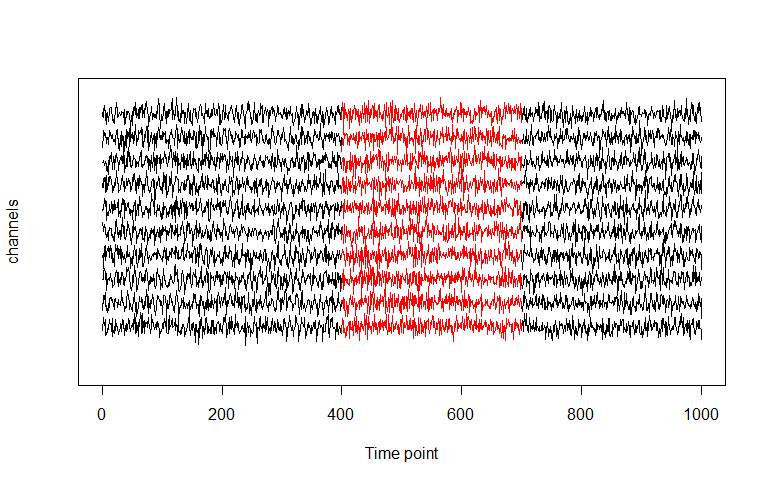}&\includegraphics[scale=0.35]{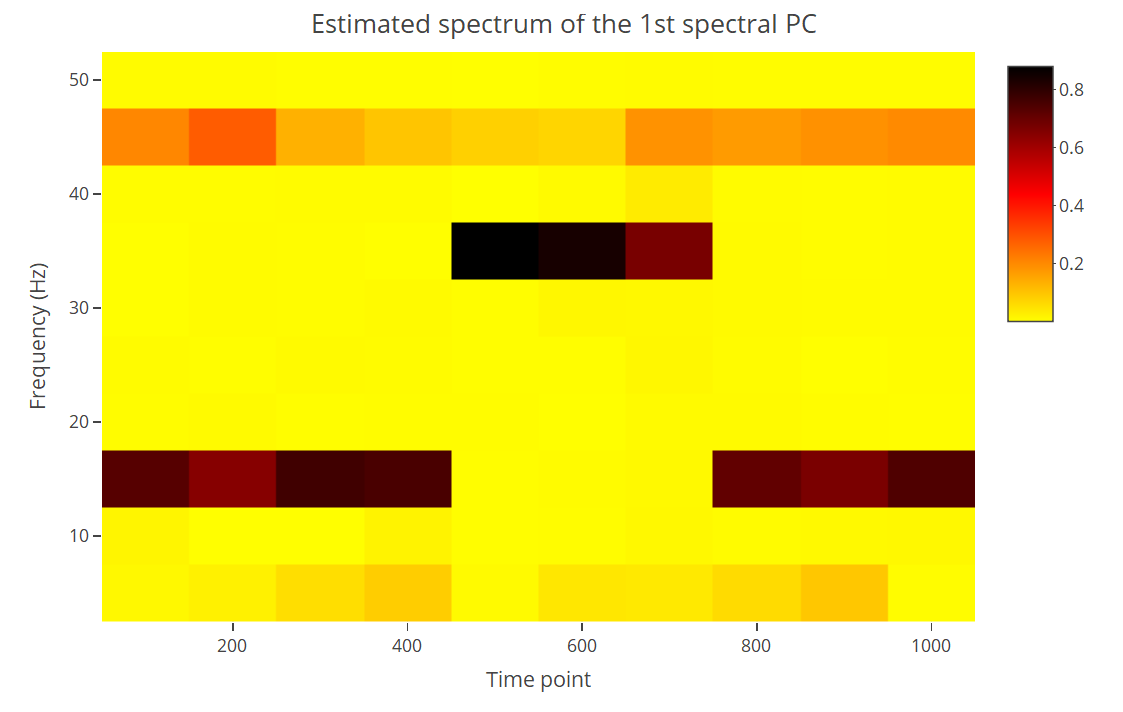}\\
		\includegraphics[scale=0.5]{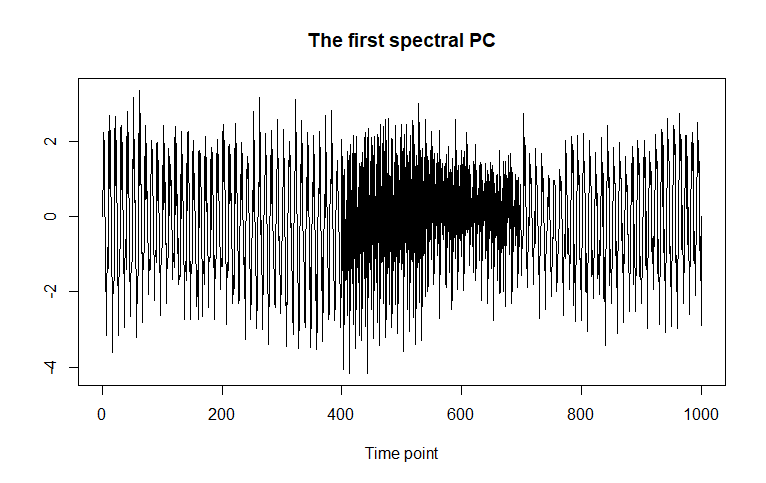}&\includegraphics[scale=0.5]{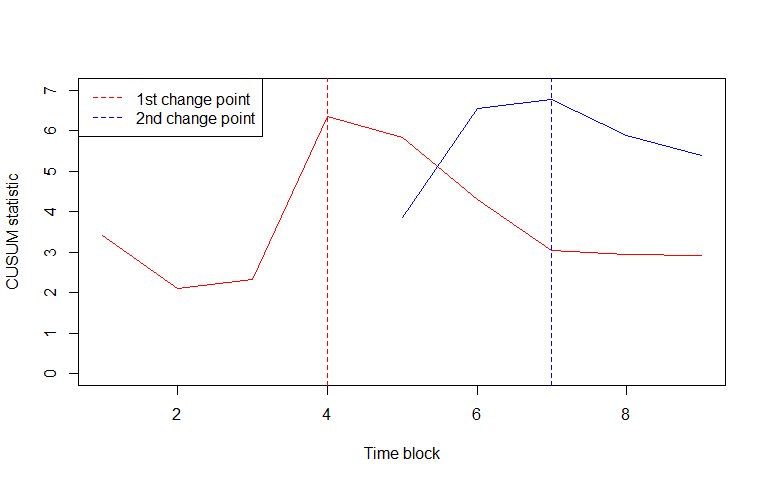}
	\end{tabular}
	\caption{Topleft: Time series with $T=1,000$ and $p=20$ channels, two change points at $t=400$ and $700$. Topright: The time-frequency plot. Presented are estimated spectrum of the first spectral PC at each time block $(blocklength = 100)$. Bottomleft: The first spectral PC and estimated change points. Bottomright: CUSUM statistics using binary segmentation (Red line: first change point. Blue line: second change point.)}
\end{figure}
\FloatBarrier
Suppose that the entire time has been split into $\tilde{T}$ blocks. 
The CUSUM statistic on the $\tilde{t}$-th block is defined as 
\begin{equation}
\centering
\mathcal{C}_{\tilde{t}} = \sum\limits_{\omega_j}\mathcal{C}_{\tilde{t}}(\omega_j)*\mathbb{I}(\mathcal{C}_{\tilde{t}}(\omega_j)>\tau_T)
\end{equation}
where 
\begin{equation}
\centering
\mathcal{C}_{\tilde{t}}(\omega_j) = |\sqrt{\frac{\tilde{T}-\tilde{t}}{\tilde{T}\tilde{t}}}\sum\limits_{i=1}^{\tilde{t}}\hat{f}(i,\omega_j)-\sqrt{\frac{\tilde{t}}{\tilde{T}(\tilde{T}-\tilde{t})}}\sum\limits_{i=\tilde{t}+1}^{\tilde{T}}\hat{f}(i,\omega_j)|/\hat{\sigma}(\hat{f}(\omega_j))
\end{equation}
is the CUSUM statistic at frequency $\omega_j$. Here, $\tau_T = 0.8\log_{1.1}(T)$ is a threshold, whose the theoretical justification has been provided by Schr\"oder and Ombao (2019). 
$\hat{\sigma}(\omega_j) = \frac{1}{\tilde{T}}\sum_{l=1}^{\tilde{T}}\hat{f}(l,\omega_j)$ is a scaling factor. 

In the binary segmentation CUSUM algorithm, for each time block we calculate the periodogram series $\hat{f}(\tilde{t},\omega_j)$ at all frequency $\omega_j$ based on a spectral PC $u_\ell(t)$. Then we identify the first candidate change-point that gives the largest discrepancy of CUSUM statistics. Once a change point is detected at time block $t_0$ $(1<t_0<\tilde T)$, we repeat our procedure on sub-intervals split by $t_0$. The following table provides the pseudo code of our algorithm: 

\begin{algorithm}[h]
	\caption{Binary segmentation CUSUM}\label{euclid}
	\textbf{Input:} Periodogram series $\hat{f}(\omega_j) = (\hat{f}(1,\omega_j),...,\hat{f}(\tilde{T},\omega_j))$, start and end point of the series, threshold $\tau_T$.
	\begin{algorithmic}[1]
		\Function {Binary-CUSUM($\hat{f}(\omega_j)$,$start$,$end$,$\tau_T$)}{}
		\If {$end-start\ge 2$} 
		\For{Block $\tilde{t}=1,2,...,\tilde{T}$} 
		\State Calculate $\mathcal{C}_{\tilde{t}}$ based on $\hat{f}(start, \omega_j),..., \hat{f}(end, \omega_j)$ using $(1),(2)$
		\EndFor
		\State $t_0=\arg\max_{\tilde{t}\in[start,end]}\mathcal{C}_{\tilde{t}}$
		\If {$\mathcal{C}_{t_0}> \tau_T$ 
		}
		\State Add $t_0$ to the set of estimated change points.
		\State Binary-CUSUM($\hat{f}(\omega_j)$,$start$,$t_0$,$\tau_T$).
		\State Binary-CUSUM($\hat{f}(\omega_j)$,$t_0+1$,$end$,$\tau_T$).
		\Else 
		\State{Break.}
		\EndIf
		\EndIf
		\EndFunction
	\end{algorithmic}
\end{algorithm}
\FloatBarrier
Our two-stage Spec PC-CP method is summarized in Figure 2. In the first stage, we partition the whole time series into time blocks of length $100$ and compute the first spectral PC at each time interval (Figure 2(b)). In the second stage we compute the periodogram at each time block using the PCs (Figure 2(c)) and conduct change point detection using the binary segmentation CUSUM procedure (Figure 2(d)).
\FloatBarrier

\subsection{Other methods for comparison}
Besides Contemporaneous Mixture Method,
we also compared our method with two other existing methods, namely Structural Break Detection (Safikhani and Shojaje, 2017) and Sparsified Binary Segmentation (Cho and Fryzlewicz, 2015). Both of the methods aim to detect change points in multivariate time series.
The Structural Break Detection method is a parametric method based on a piecewise vector autoregressive model (VAR) with a regularized estimation using the total variation LASSO penalty.
The Sparsified Binary Segmentation method is a non-parametric method that aggregates the CUSUM statistics with a sparsifying step that reduces the effect of noisy contributions. To reduce the impact of the channels with no changes, it applies a threshold to the CUSUM statistics for each single channel and aggregates only the CUSUMs that exceeds the threshold. 

\section{Simulation Studies}
	\subsection{Generate the latent sources}
In this section, we present simulation studies to examine the performance of our proposed Spec PC-CP method and three existing methods. We consider several methods for generating latent source. 
To obtain realistic signals, we follow the approach in Gao et al. (2016) to generate latent sources using the AR(2) model. As demonstrated in Gao et al. (2016), the AR(2) model is able to produce signals with the desired oscillatory properties. Following the convention for analyzing electroencephalograms (EEGs), we generated sources with localized power in the following frequency bands: delta (0-4) Hz,
theta (4-8) Hz, alpha(8-12) Hz, beta(12-30) Hz, gamma(30-45) Hz. The synthetic signals are generated as follows:
\begin{itemize}
	\item We first generate multivariate AR(2) latent sources using $\mathbf{S}(t)=\Phi_1 \mathbf{S}({t-1})+ \Phi_2 \mathbf{S}(t-2)+\mathbf{\eta}(t)$, where $\Phi_1, \Phi_2\in \mathbb{R}^{k\times k}$  are diagonal matrices and noise $\eta(t)$ is independent Gaussian to guarantee independence of the sources, $\mathbf{\eta}(t)$ are Gaussian noise and each column follows $N(\mathbf{0_{k\times 1}}, \mathbf{I}_k)$. Note that we consider five (k = 5) different latent AR(2) processes that represent different EEG oscillations at the five frequency bands respectively. The coefficients for each of the AR(2) component are listed in table 1.
	\begin{center}
		\captionof{table}{AR(2) Coefficients for each frequency band}
		\begin{tabular}{c c c }
			\hline
			Frequency band  & Diagonal value of $\Phi_1$  &  Diagonal value of $\Phi_2$  \\
			
			\hline
			Delta   &1.998 &-0.998  \\
			Theta   &1.9 &-0.998    \\
			Alpha   &1.616 &-0.998    \\
			Beta   &-0.617 &-0.998    \\
			Gamma   &-1.616 &-0.998   \\
			\hline
		\end{tabular}
	\end{center}

	\item We then generate the $p$-dimensional signals from $k$ latent sources as follows: 
	\begin{equation*}
	X(t)=M\mathbf{S}(t)+\epsilon(t)
	\end{equation*}
	Each column of the noise term $\epsilon(t)$ follows $N(\mathbf{0_{p\times 1}}, \mathbf{I}_p)$. In the simulation study, we are interested in multiple factors that might affect the results, such as different fractions of channels with changes, the number of change points, different types of latent source, the change point locations and different block lengths. Figure 3 shows the latent source in theta band and alpha band. The sampling rate is 100 Hz.  By using AR(2) model, we can constrain the power of sources to center at given frequency bands.
\end{itemize} 

\begin{figure}[h]
	\centering
	\begin{tabular}{c c }
		\includegraphics[scale=0.3]{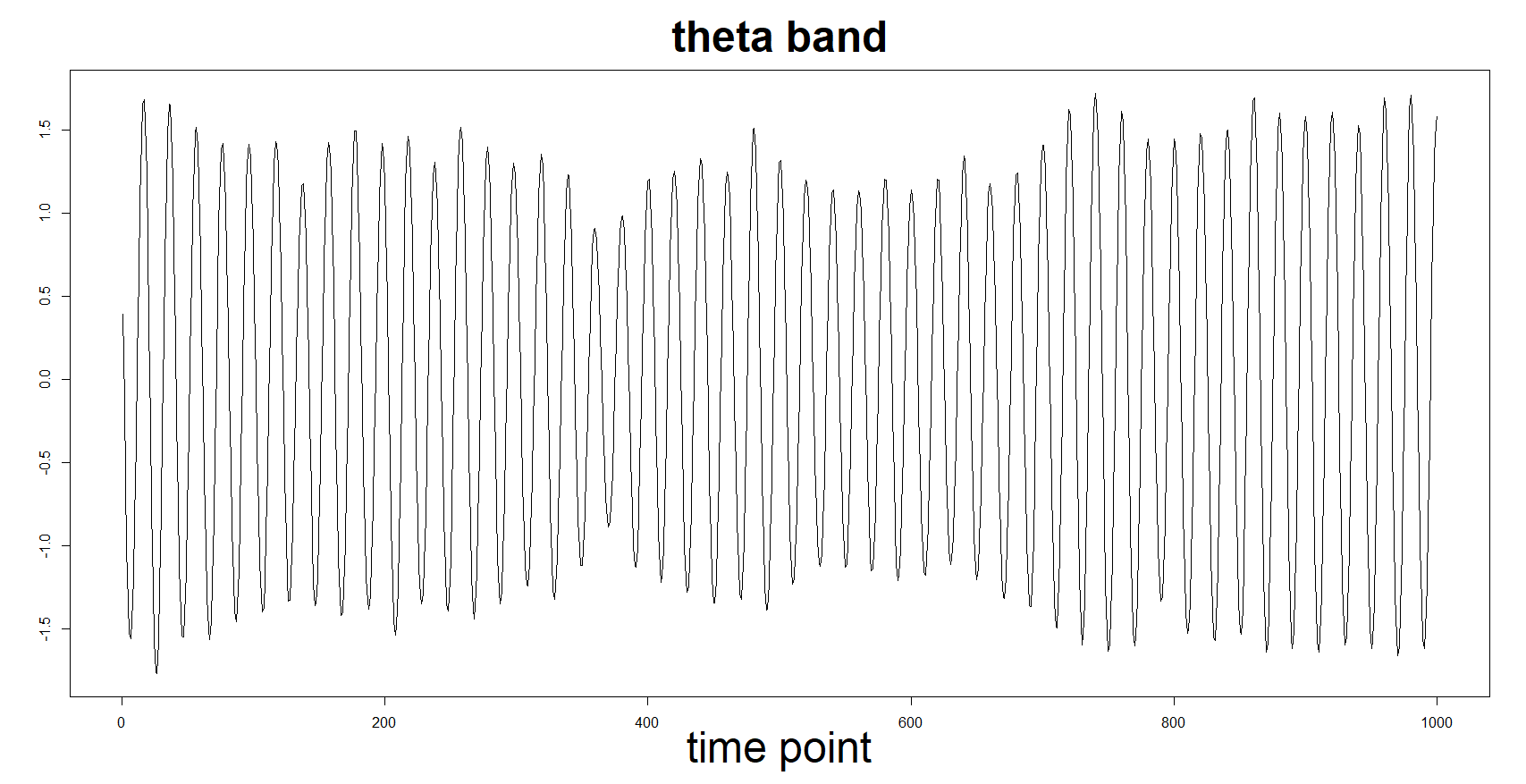}&\includegraphics[scale=0.3]{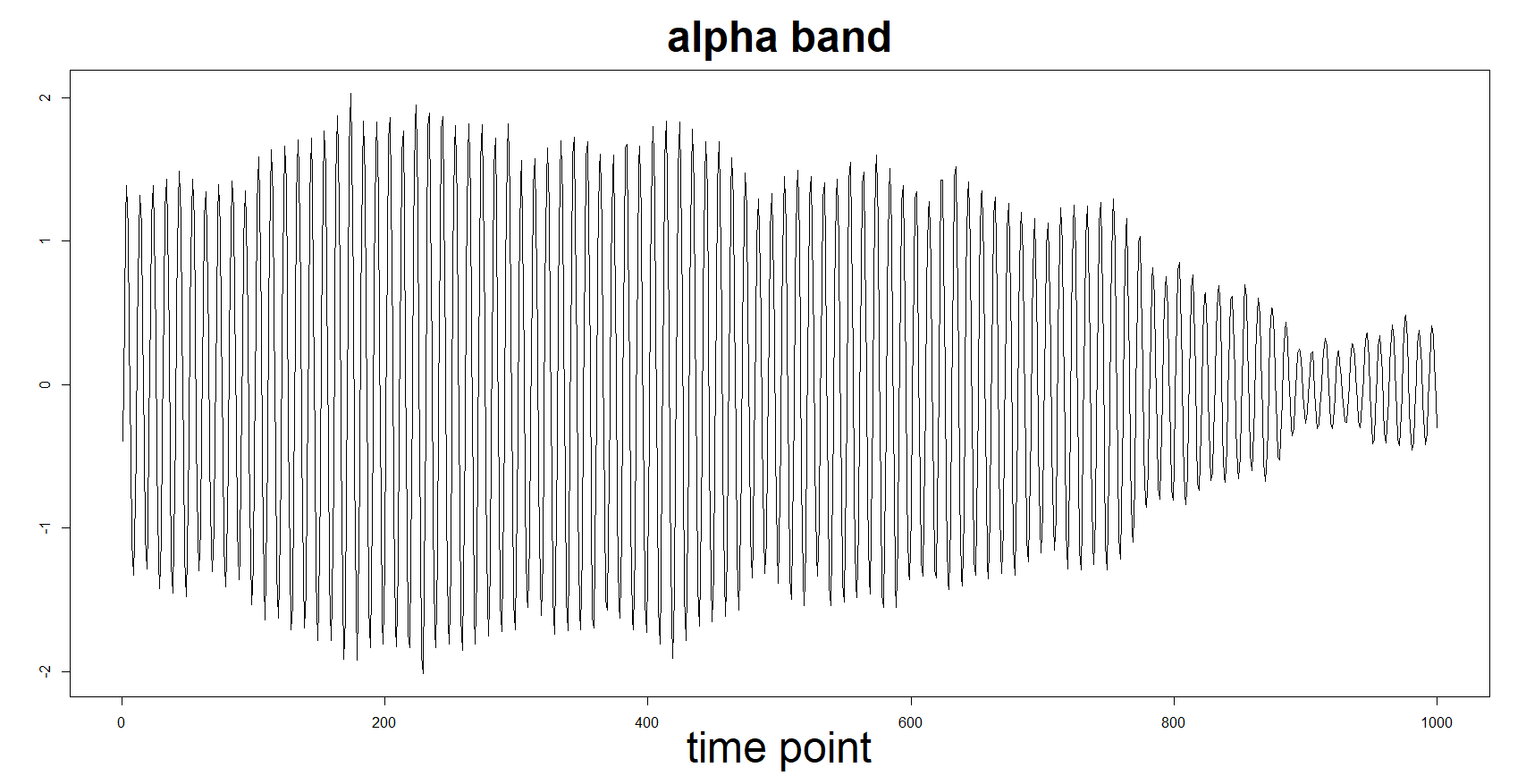}\\
		\includegraphics[scale=0.7]{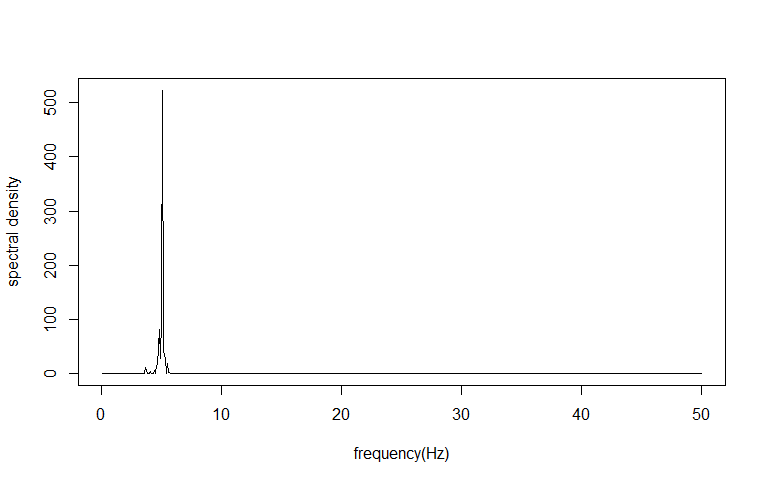}&\includegraphics[scale=0.7]{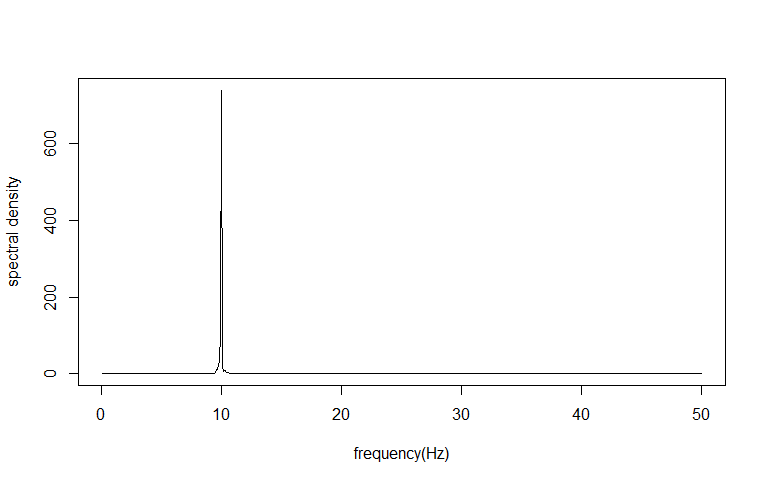}\\
	\end{tabular}
	\caption{The series and spectra of the process in theta band and alpha band. Sampling rate: 100 Hz. (Top: series, bottom: spectra)}
\end{figure}
\FloatBarrier

\subsection*{Summary of Simulation Settings}
To provide sensitivity analysis on our simulation, we used a variaty of simulations settings, which are summarized as follows:
\begin{itemize}
	\item The length of time series is $T=1,000$ with 1 change point(at $t=550$) or $T=4,000$ with 3 change points(at $t=980;2,120;3,050$).
	\item The number of channels with change points (out of total of $p$=128) is 16, 32, 64.
	
	\item The type of latent source is AR(2) or a mixed source of ARMA(1,1), MA(1) and AR(2).
	\item The first three spectral principal components that correspond to the first three largest eigenvalues from Spectral PCA.
\end{itemize}

\subsection{Scenario I: Proportions of channels with a change point }
First we evaluate the case where we have AR(2) latent source, 1 change point at $t=550$ and $T=1,000$. We assume that changes mainly in the gamma band(the $5$-th column of the weight matrix) and add time lag with length 10 or 15 on $X_t$ to create lead-lag relationship. Under such settings we expect the Spectral PCA method could better summarize the time series because of its ability to extract the latent source and capture the dynamics of the source with time lag. A total of $p=128$ channels will be generated. The following numbers of channels with change points are considered: 16, 32, 64. \\

\[
M_{10}^{(1)} =\begin{pmatrix}
0.1    & 0.1 &  0.1&0.1 & 0.1 \\
0.1    & 0.1 &  0.1&0.1 & 0.1 \\
&  &  \vdots &    \\
0.1    & 0.1 &  0.1&0.1 & 0.1 \\
\end{pmatrix}
\quad
M_{15}^{(1)} =\begin{pmatrix}
0.2    & 0.1 &  0.1&0.1 & 0.1 \\
0.2    & 0.1 &  0.1&0.1 & 0.1 \\
&  &  \vdots &    \\
0.2    & 0.1 &  0.1&0.1 & 0.1 \\
\end{pmatrix}
\quad
\]

\[
M_{10}^{(2)} =\begin{pmatrix}
0.1 &  0.1&0.1 & 0.1 & 0.9 \\
0.1 &  0.1&0.1 & 0.1 & 0.9 \\
&  &  \vdots &    \\
\end{pmatrix}
\quad
M_{15}^{(2)} =\begin{pmatrix}
0.3    &  0.1&0.1 & 0.1  & 0.9 \\
0.3    &  0.1&0.1 & 0.1  & 0.9 \\
&  &  \vdots &    \\
\end{pmatrix}
\quad
\]

\[ X_t = \left\{
\begin{array}{ll}
M_{10}^{(1)}S(t-10)+M_{15}^{(1)}S({t-15})+\epsilon(t) & (0<t<=550)\\
M_{10}^{(2)}S({t-10})+M_{15}^{(2)}S({t-15})+\epsilon(t) & (550<t<=1,000)
\end{array}
\right.
\]
\FloatBarrier
where each column of $\epsilon(t)$ follows $N(0_{k\times 1},I_k)$.\\

\begin{figure}[h]
	\centering
	\begin{tabular}{c c }
		\includegraphics[scale=0.5]{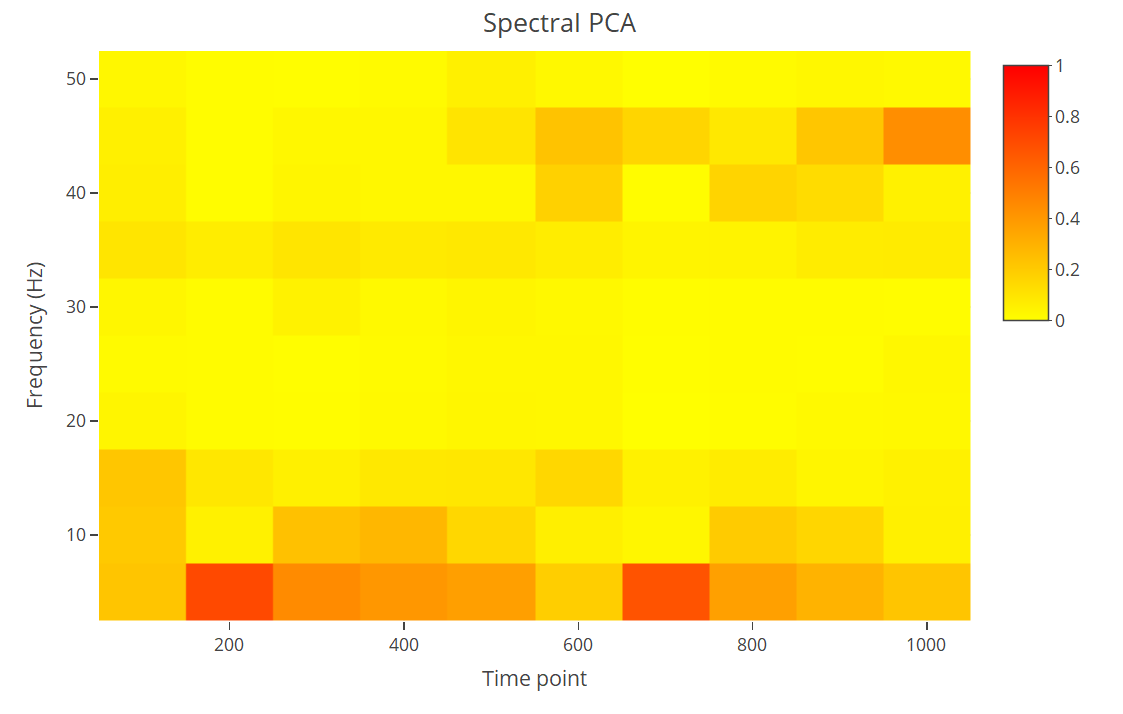}&\includegraphics[scale=0.5]{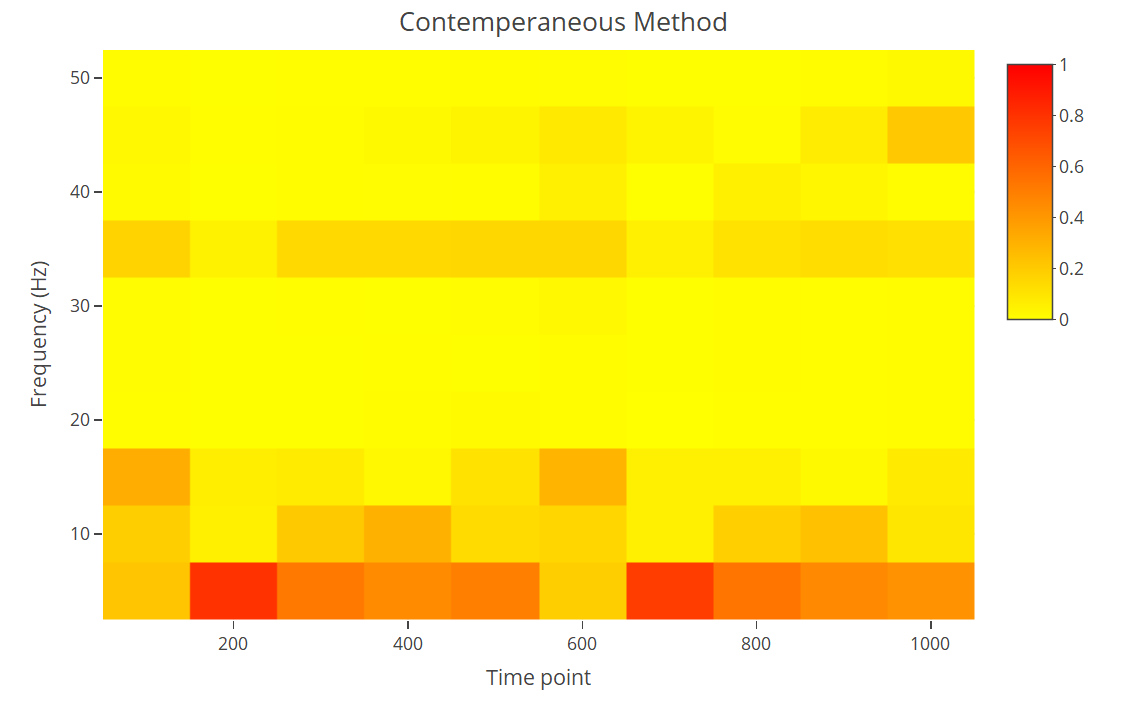}\\
		\includegraphics[scale=0.5]{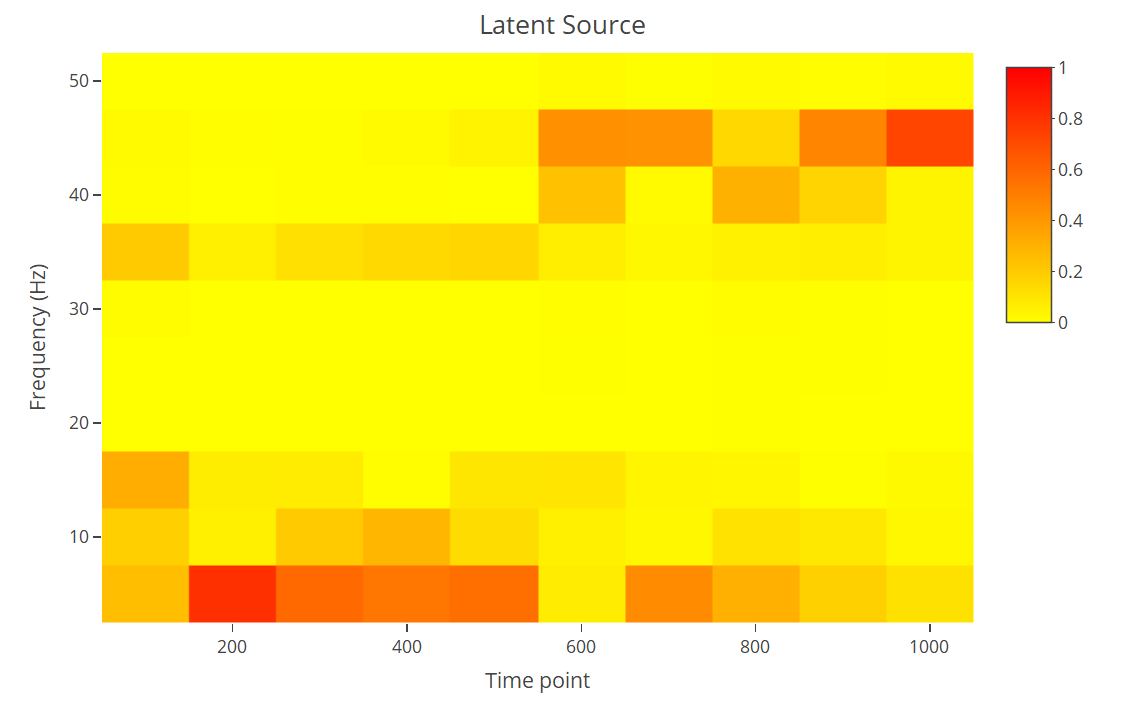}\\
	\end{tabular}
	\caption{The Time-Frequency plots of the summarization series(top-left: the first Spectral PC, top right:  the first component of Contemporaneous  Mixture, bottom: the latent source. Total time points $T=1,000$, 16/128 channels with change point)}
\end{figure}
\FloatBarrier
As the Time-Frequency plots shown in figure 3, Spectral PCA has higher weights in high frequencies after the change point and lower weights in other frequency bands. As a comparison, the contemporaneous method has higher weights in lower frequencies. This indicates that Spectral PCA gives a better summary than Contemporaneous Method..

Table 2 gives the simulation results. Among the three leading spectral PCs, the first PC performs.
the best.
As the number of channels with change increases, the detection rate increases in all methods. Spec PC-CP and Contemporaneous Method have comparable performance when there are a large number of channels with change points. The advantage of our Spec PC-CP method is particularly substantial when the fraction of channels with change points is small (16 channels). Besides the detection rate, our method is more accurate in locating the change point, as qualified in mean absolute difference (MAD) between the estimated change points and true ones (Figure 4 and Table 2). The change points detected by Spec PC-CP are closer to the true change points ($t=550$) than contemporaneous method, especially when there is a small proportion of channels with change points. To confirm the robustness of our method against simulation methods, we also run simulations under Structural Break and Binary Segmentation settings. Table 5 and 6 in appendix show the results. In these settings there is one change point and the Spec PC-CP method gives comparable results in terms of detection rate and slightly higher mean absolute difference.

\subsection{Scenario II: Multiple Change Points}
Next we conduct simulations with multiple change points. The length of time series lengths are chosen as $T=4,000$, and the three change points are at $t=980;2,150;3,020$.

\[ X(t) = \left\{
\begin{array}{ll}
M_{10}^{(1)}S({t-10})+M_{15}^{(1)}S({t-15})+\epsilon(t) & (0<t<=980)\\
M_{10}^{(2)}S({t-10})+M_{15}^{(2)}S({t-15})+\epsilon(t) & (981<t<=2,150)\\
M_{10}^{(1)}S({t-10})+M_{15}^{(1)}S({t-15})+\epsilon(t) & (2,151<t<=3,020)\\
M_{10}^{(2)}S({t-10})+M_{15}^{(2)}S({t-15})+\epsilon(t) & (3,021<t<=4,000)
\end{array}
\right.
\]
\FloatBarrier
where each column of $\epsilon(t)$ follows $N(0_{k\times 1},I_k)$.\\

The performance of the four methods are summarized in Table 3. Again, the Spec PC-CP method performs the best. In particular, our method outperforms the other methods when there are low or moderate number of channels (16 and 32). When the true number of change point is one, we used detection rate as one metric to compare the performance of different methods. Since there are multiple change points in this setting, it is more reasonable to compare the detection proportion, which is the number of detected change points over total number of change points. The Structural Break Method has about 0.33 detection proportion and only detects 1 change point at the very end of the series in most cases. The Sparsified Binary Segmentation gives lower detection proportion around 0.4 with higher bias. Figure 5 shows the estimated locations for Spec PC-CP and contemporaneous method. Change points detected by our Spec PC-CP are closer to the true change points. Moreover, most of the estimated change points are in a small neighborhood of the true locations ($t=1,000; 2,100; 3,100$). 

\subsection{Scenario III: Mixed sources}
In this setting, we considered mixed latent sources with ARMA(1,1), MA(1) and AR(2) sources.  

\[ X(t) = \left\{
\begin{array}{ll}
M_{10}^{(1)}S({t-10})+M_{15}^{(1)}S({t-15})+\epsilon(t) & (0<t<=550)\\
M_{10}^{(2)}S({t-10})+M_{15}^{(2)}S({t-15})+\epsilon(t) & (550<t<=1,000)
\end{array}
\right.
\]
where each column of $\epsilon(t)$ follows $N(0_{k\times 1},I_k)$. Here $S(t)$ is a mixed latent source with ARMA(1,1), MA(1) and AR(2). The series lengths are $T=1,000$ and there is one change point at $t=550$.

The results are summarized in Table 4 and Figure 6. The spec PC-CP method using the first or second spectral PC outperform contemporaneous and structural break method when there are less channels (16) with change points and gives comparable results when mthe number of channels with change points is moderate or large. Our method has detection rates mostly from 0.6 to 0.9 in this case.  Although structural break method has moderate detection rates as well, it tends to overestimate the number of change points and has inferior performance in terms of the estimated locations. The Sparsified Binary Segmentation has inferior performance with around 0.3 detection rate. Besides, The locations of change point detected in Figure 6 of our method are mostly near the truth, which shows that Spec PC-CP also outperforms contemporaneous method under this setting. The results indicates that our method is robust with different type of sources.

\begin{sidewaystable}[h!]
	\caption{Summary of simulation results for the AR(2) model.  Time series length is $T=1,000$. There is one change point at $t=550$.
		\\ Presented are the detection rates and the mean absolute distances (MADs) between the estimated and true change point.}
	\centering
	\begin{small}
		\begin{tabular}{llccccccc}
			\hline
			&& \multicolumn{2}{c}{Detection Rate} && \multicolumn{3}{c}{MAD}\\
			\cline{3-5} \cline{7-9}
			\multicolumn{2}{c}{} & 16/128   & 32/128 & 64/128 && 16/128 & 32/128 & 64/128 \\ 
			\hline
			
			1st component 	& spec PC CP 	&	\textbf{0.92}	&  \textbf{0.98}   &0.96  	&&	51.8        & 52.0	&  52.1 \\
			& Contemporaneous Method   	&      0.72        &  0.92   & 0.97   	&&    54.7 & 53.3 	&  50.9 \\

			2nd component	& spec PC CP 	&	\textbf{0.87}	&  0.90   &{0.91}  	&&	61.2        & 56.2 	&  52.2 \\
			& Contemporaneous Method   	&      0.66        &  0.90   & 0.92   	&&    59.3 	& 55.3 	&  53.1 \\

			3rd component	& spec PC CP 	&	\textbf{0.75}	&  {0.77}   &{0.84}  	&&	63.3        & 65.6 	&  60.9 \\
			& Contemporaneous Method   	&      0.65        &  0.81   & 0.92   	&&    56.4 	& 51.4 	&  53.4 \\
			\hline
			& Structual Break   	&      0.34        & 0.77   & 0.62  	&&    439.4 	& 440.1 	&  439.1 \\
			\hline
			& Sparsified Binary Segmentation   	&      0.22        & 0.29   & 0.44  	&&    39.6 	& 30.4 	&  26.4 \\
			\hline
			
		\end{tabular}
	\end{small}
	\label{table1}
\end{sidewaystable} 
\FloatBarrier

\begin{sidewaysfigure}[h!]
	\centering
	\begin{tabular}{c c c}
		\includegraphics[scale=0.33]{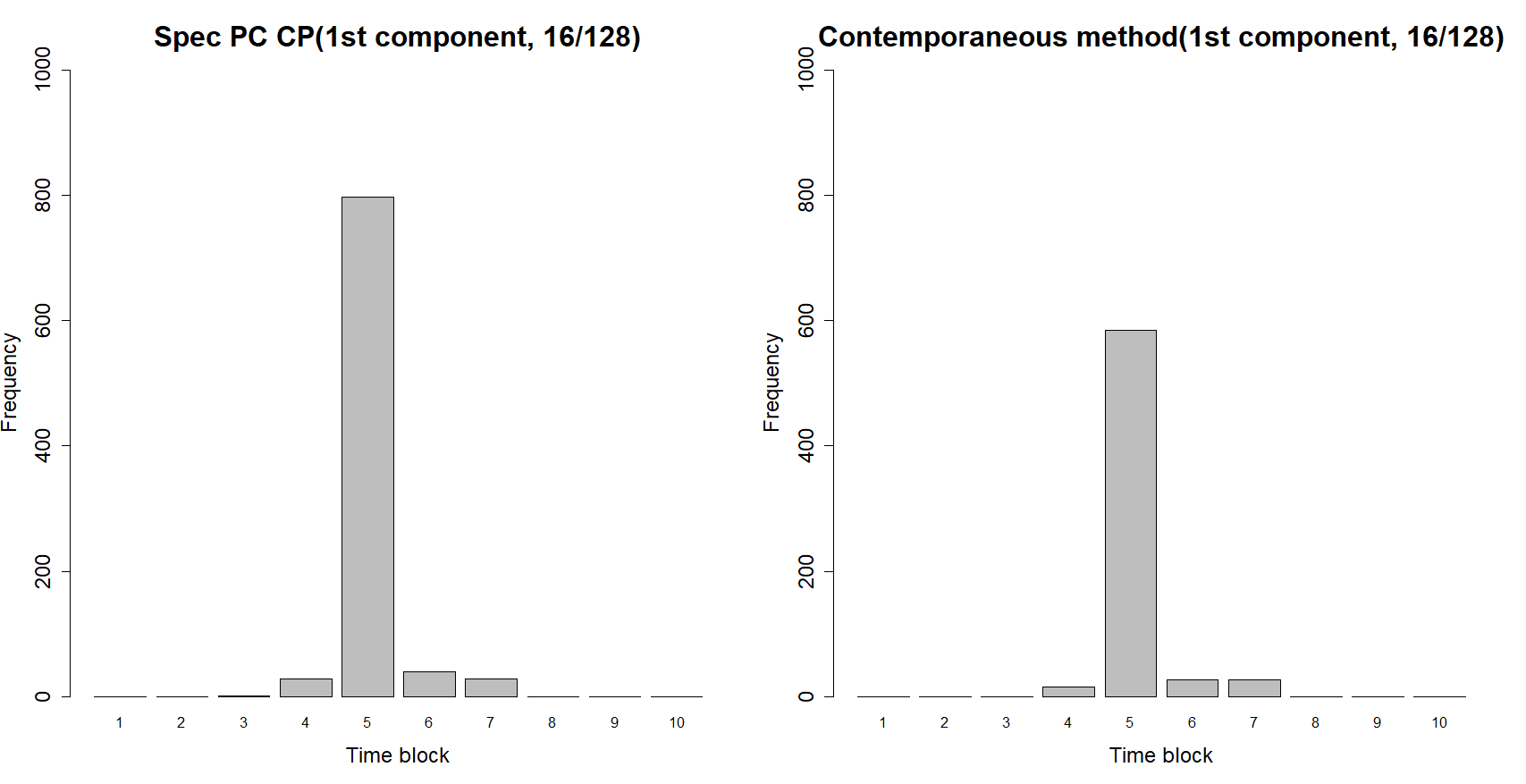}&\includegraphics[scale=0.33]{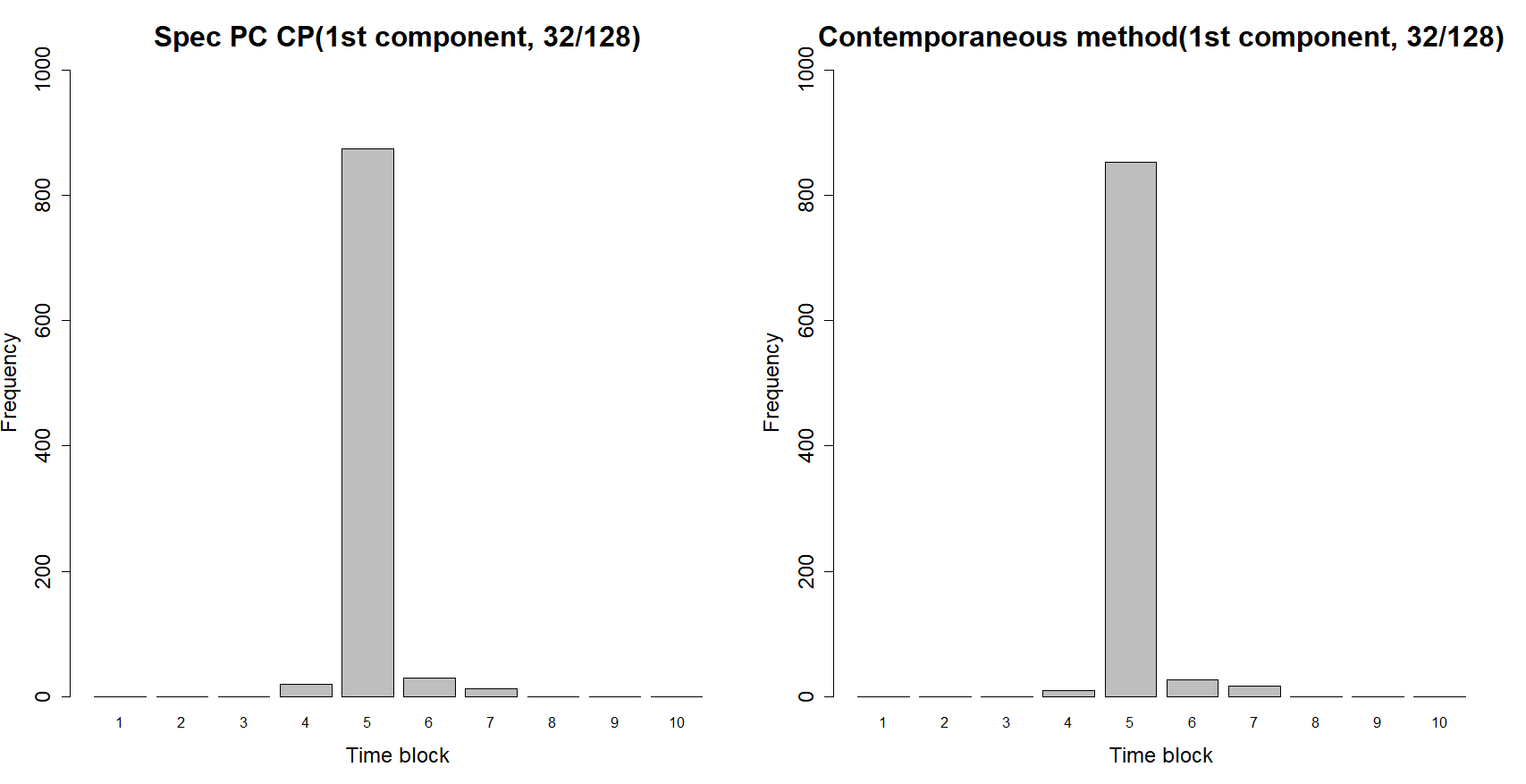}&\includegraphics[scale=0.33]{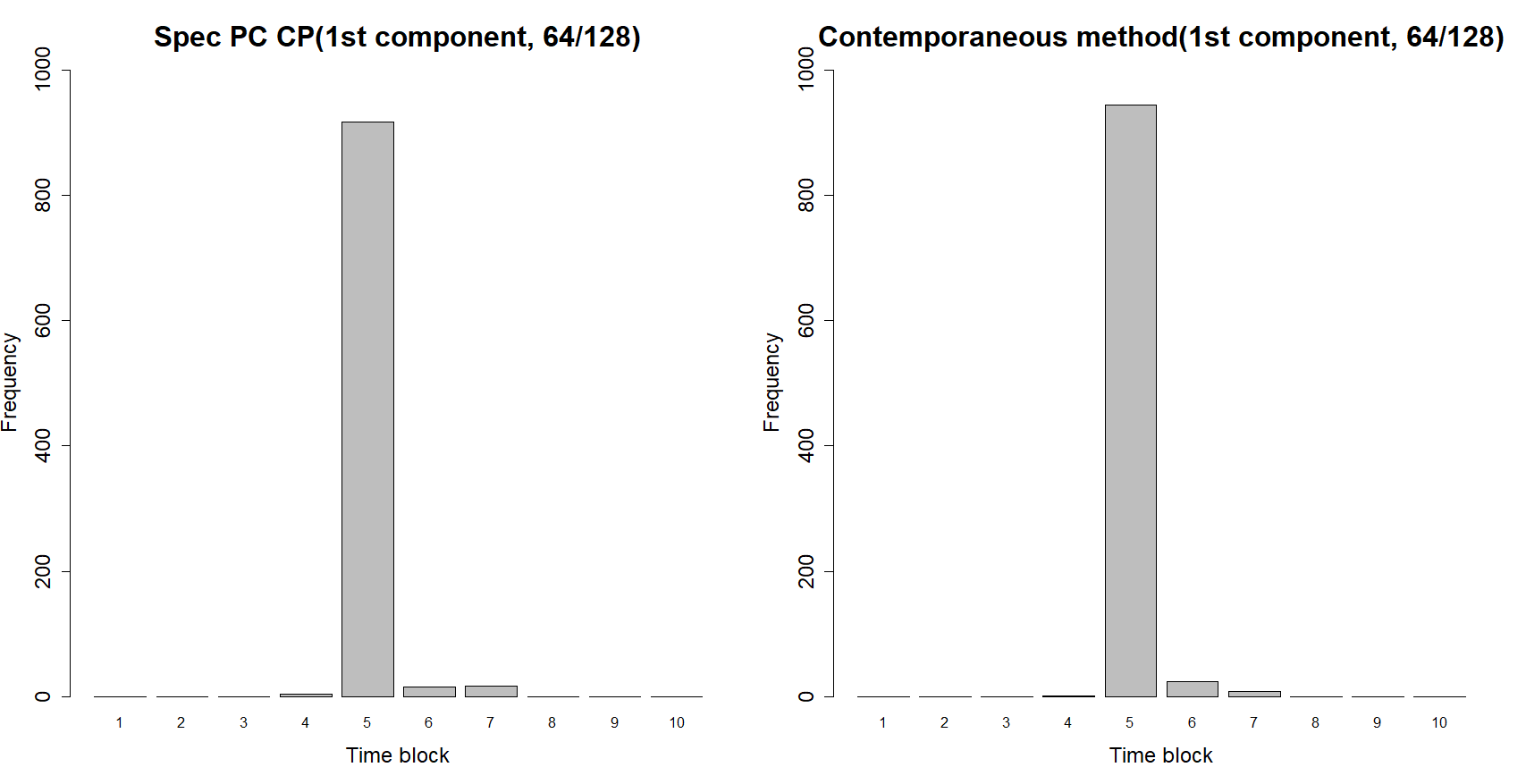}\\
		\includegraphics[scale=0.33]{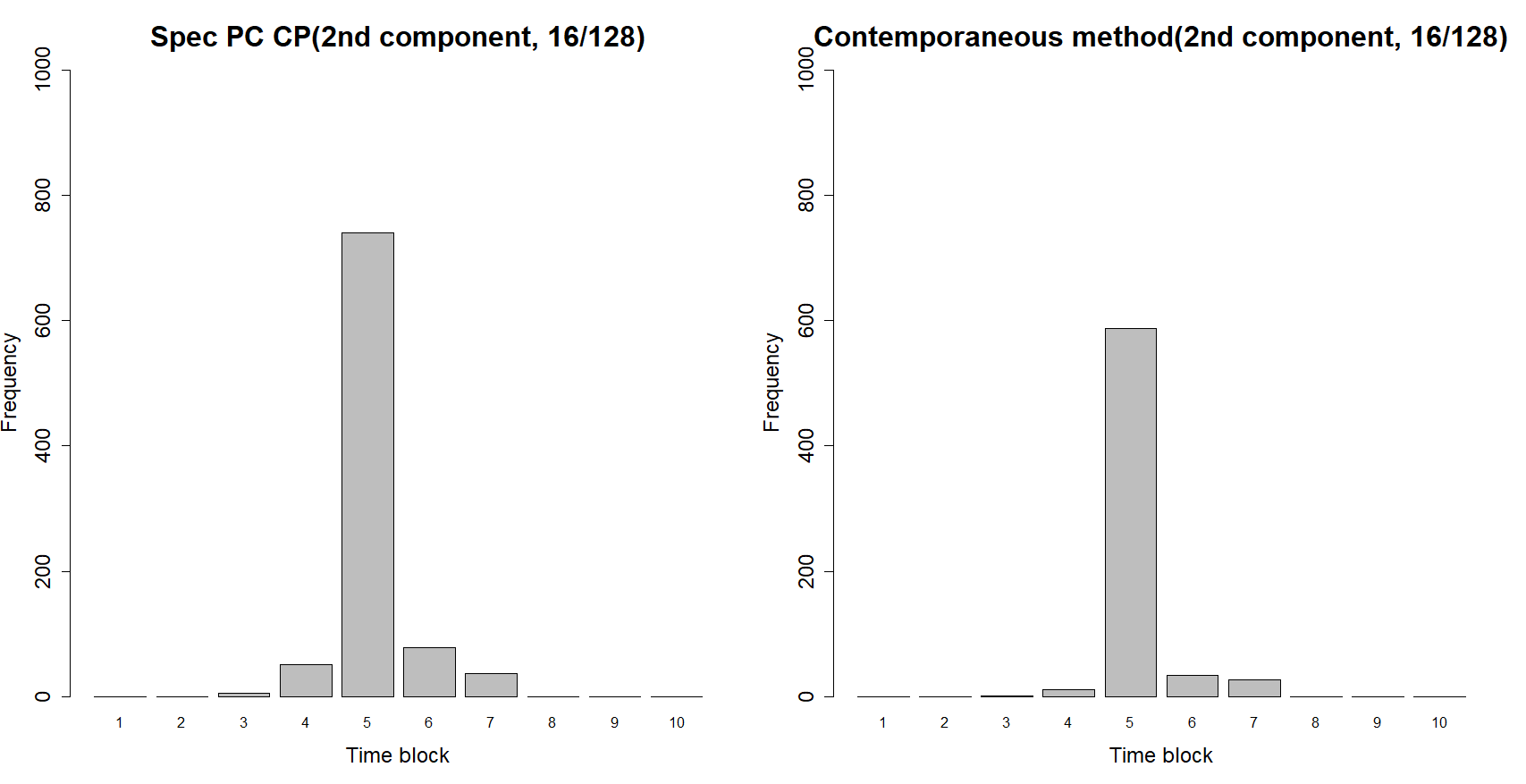}&\includegraphics[scale=0.33]{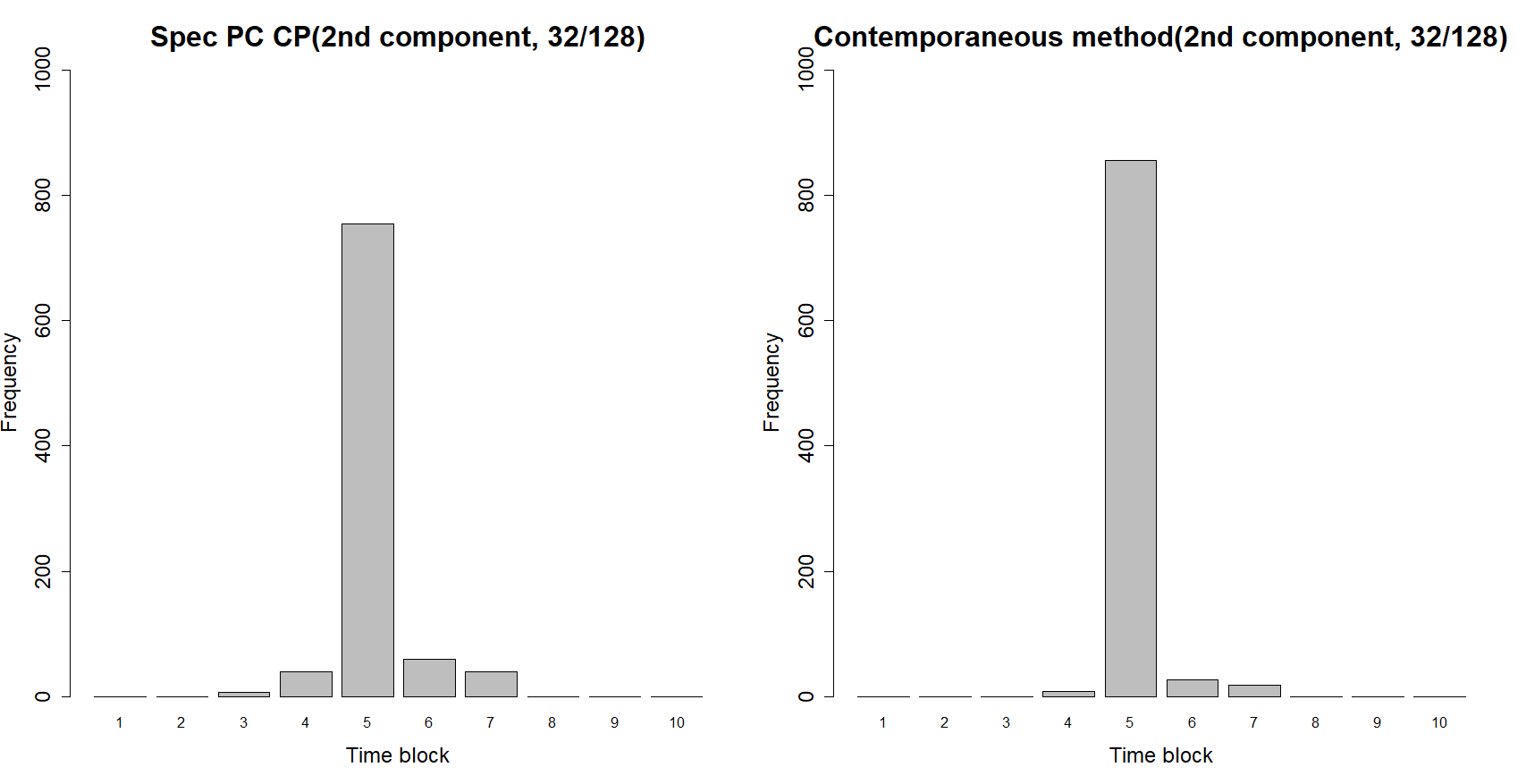}&\includegraphics[scale=0.33]{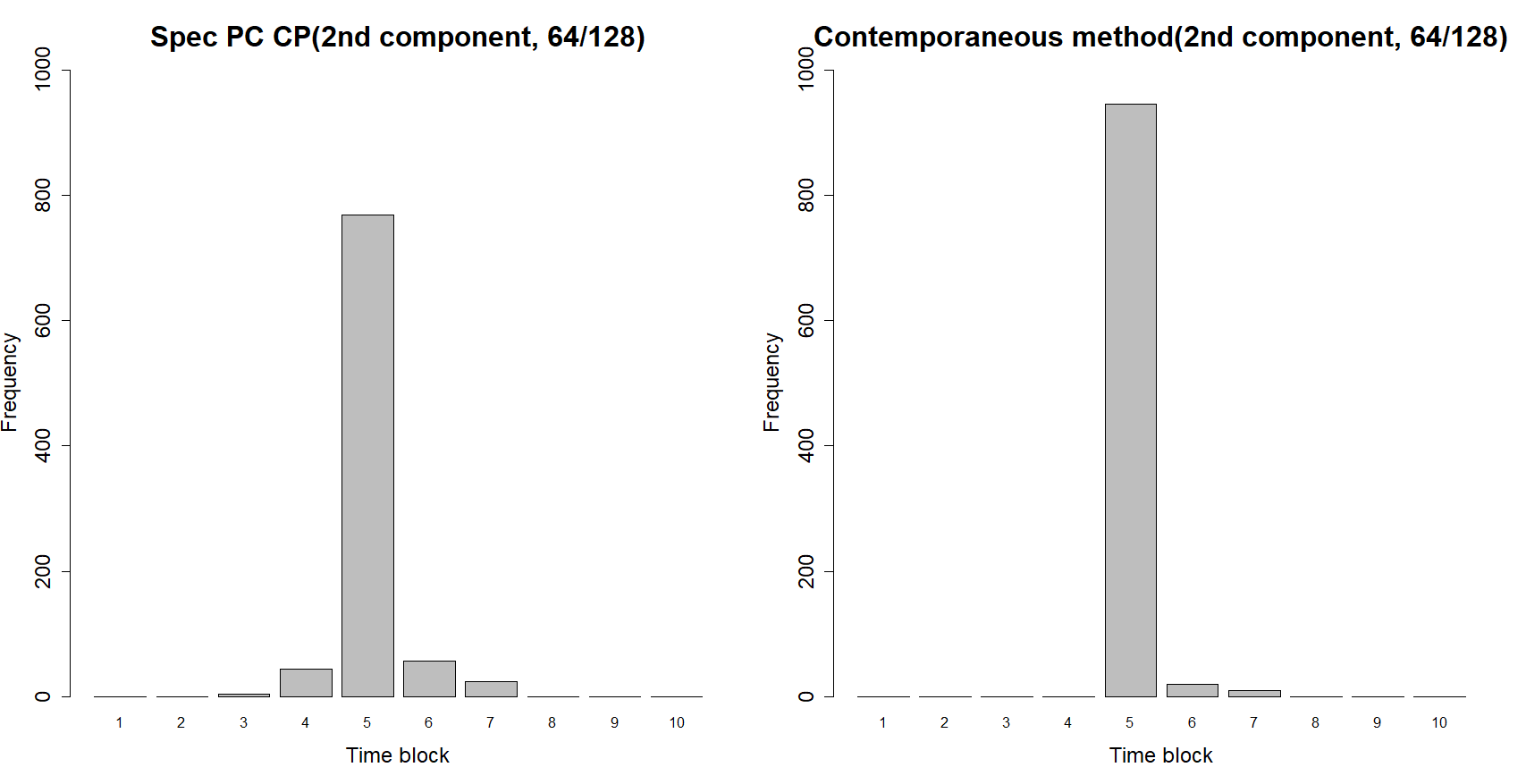}\\
		\includegraphics[scale=0.33]{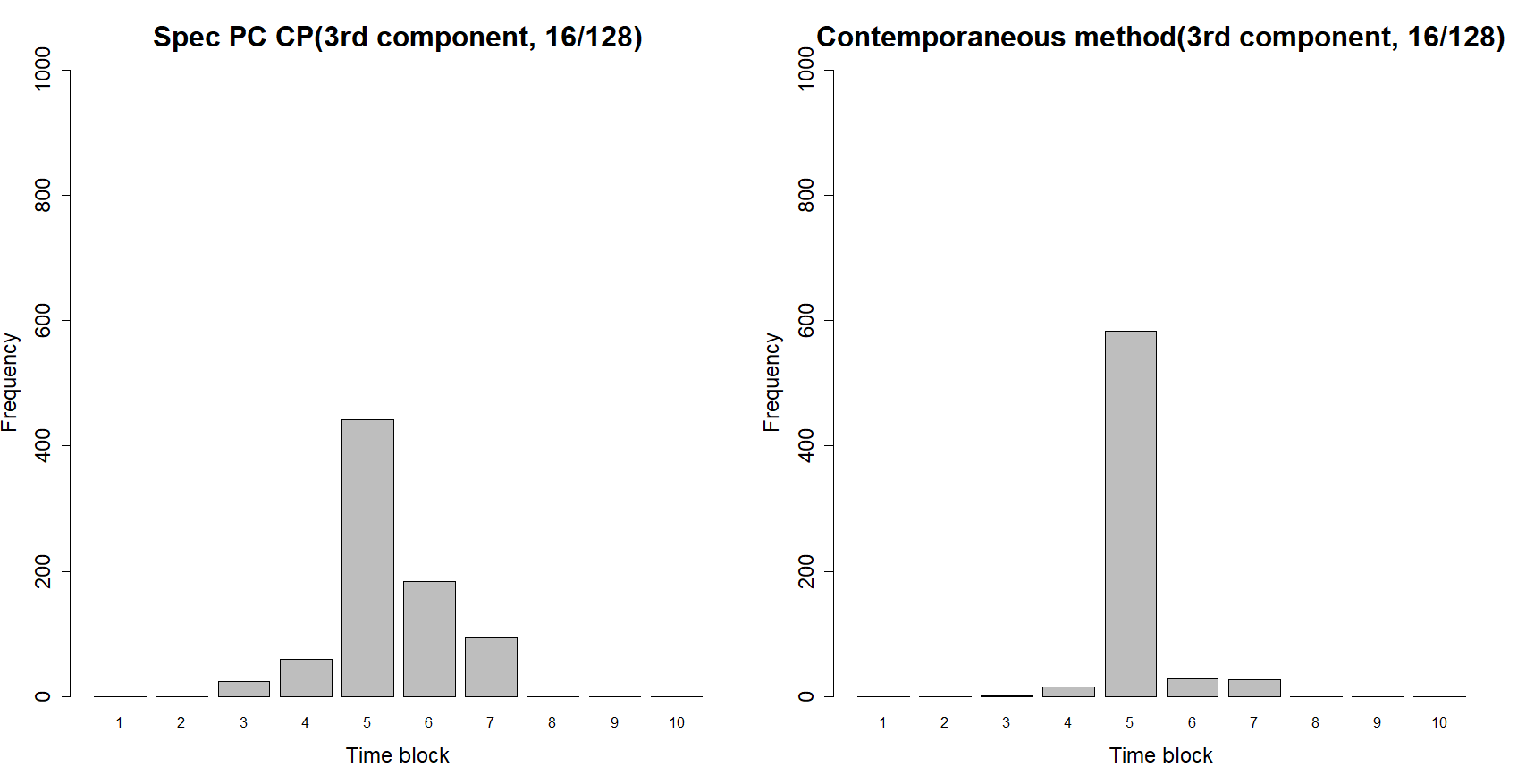}&\includegraphics[scale=0.33]{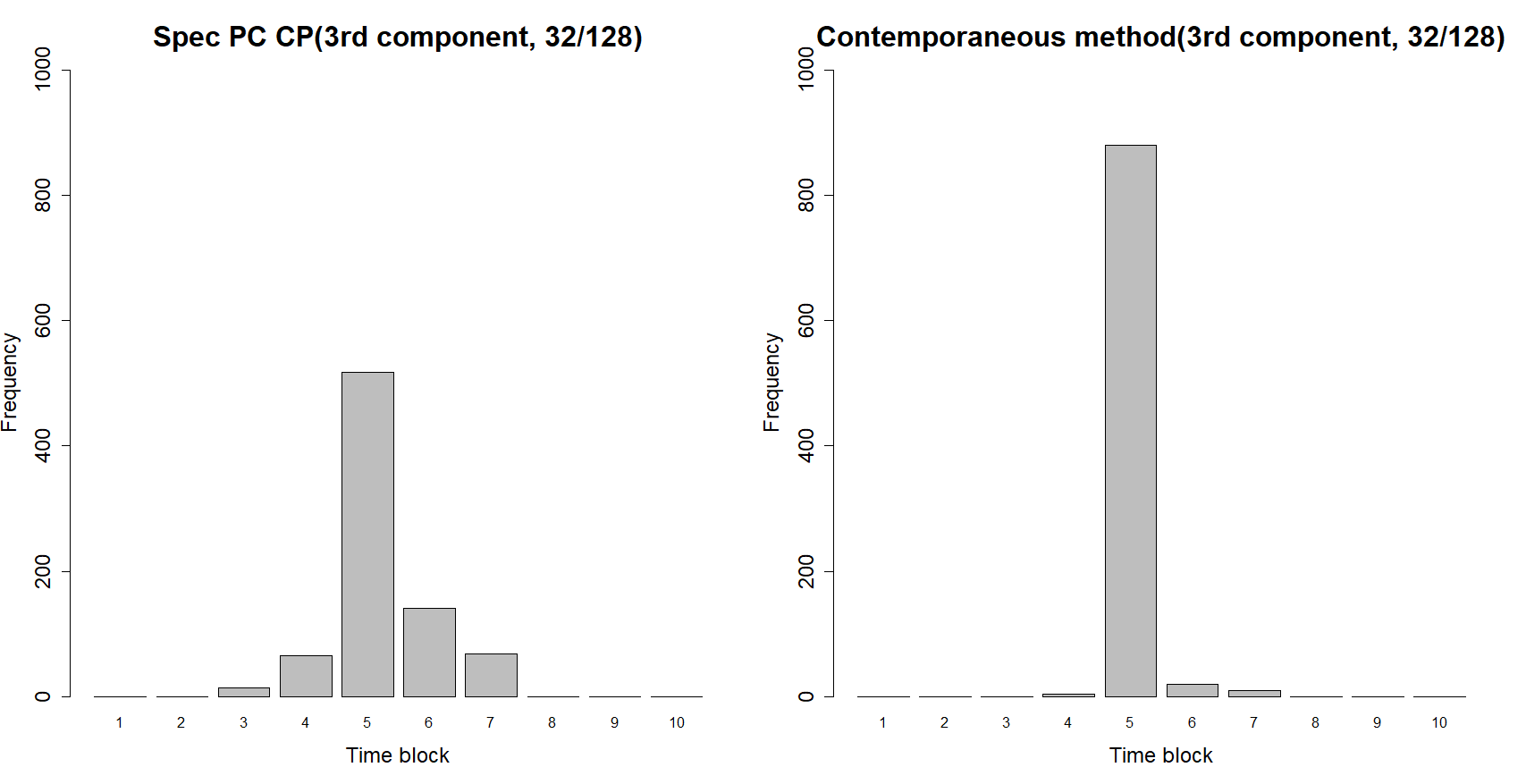}&\includegraphics[scale=0.33]{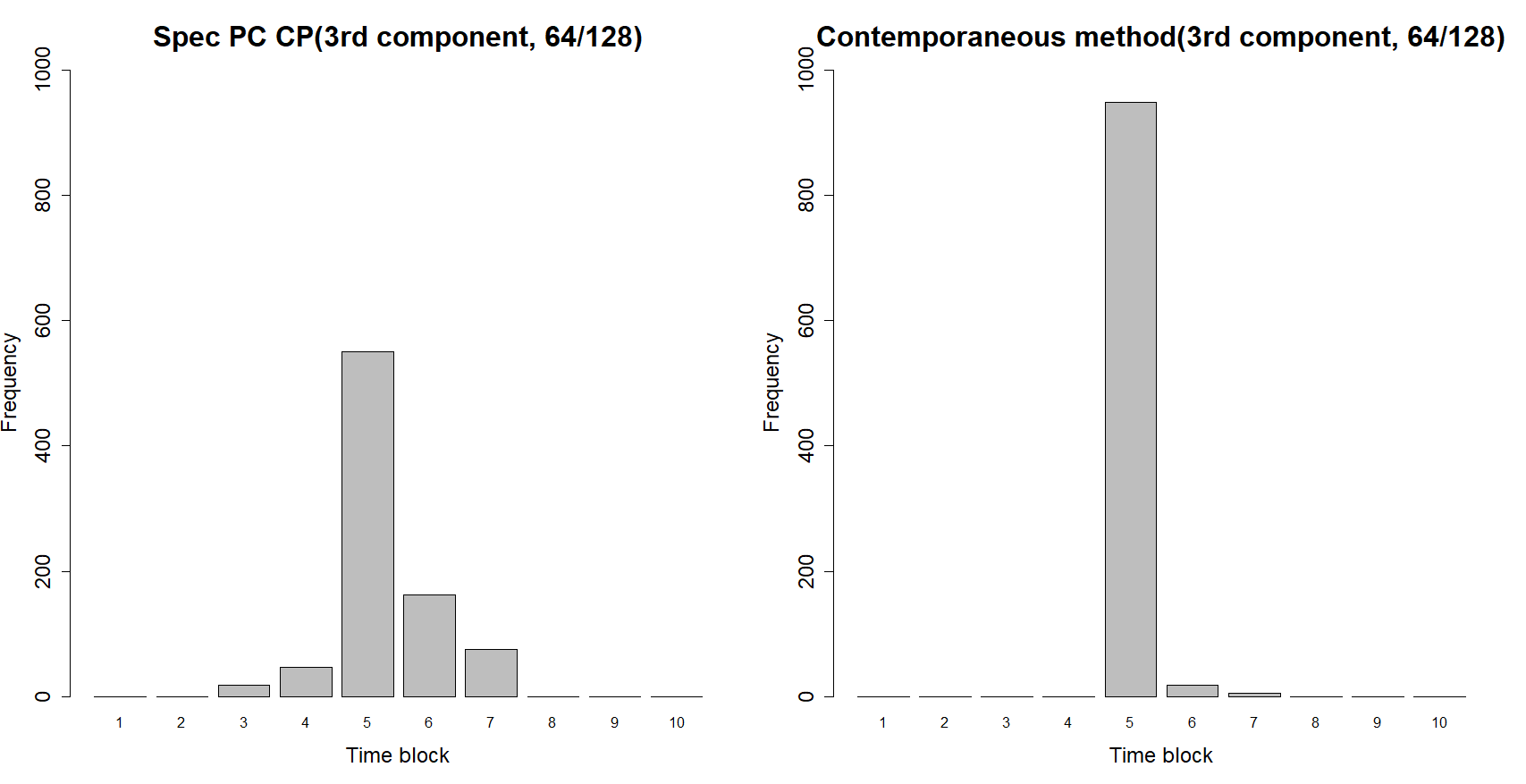}\\
	\end{tabular}
	\caption{Proportion of being detected as change point(out of 1,000). From left to right: 16,32 and 64 channels with change point, respectively. From top to bottom: using the first, second, or third spectral PC.  }
\end{sidewaysfigure}
\FloatBarrier

\begin{sidewaystable}[h!]
	\caption{Summary of simulation results for multiple change point case. Time series length is $T=4,000$. There is one change point at $t=980;2,150;3,020$.
		\\  Presented are the detection rates and the mean absolute distances (MADs) between the estimated and true change point.}
	\centering
	\begin{small}
		\begin{tabular}{llccccccc}
			\hline
			&& \multicolumn{3}{c}{Detection proportion} && \multicolumn{3}{c}{MAD}\\
			\cline{3-5} \cline{7-9}
			\multicolumn{2}{c}{} & 16/128   & 32/128 & 64/128 && 16/128 & 32/128 & 64/128 \\
			\hline
			
			1st component 	& Spec PC CP 	&	\textbf{0.71}	&  \textbf{0.84}   &0.95  	&&	885.1        & 808.3	&  321.4 \\
			& Contemporaneous Method   	&      0.46        &  0.73   & 0.96   	&&    1195.8 	& 1076.7 	&  443.5 \\

			2nd component	& Spec PC CP 	&	\textbf{0.63}	&  \textbf{0.69}   &{0.72}  	&&	544.4        & 584.5 	&  432.7 \\
			& Contemporaneous Method   	&      0.45        &  0.63   & 0.72   	&&    610.0 	& 598.1 	&  525.7 \\

			3rd component	& Spec PC CP  	&	\textbf{0.56}	&  \textbf{0.61}   &{0.66}  	&&	633.2        & 623.8 	&  346.7 \\
			& Contemporaneous Method   	&      0.45        &  0.56   & 0.64   	&&    732.4 	& 706.3 	&  413.2 \\
			\hline
			& Structural Break   	&      0.33        & 0.33   & 0.37  	&&    975.8 	& 975.7 	&  975.4 \\
			\hline
			& Sparsified Binary Segmentation   	&      0.43        & 0.42   & 0.43  	&&   1511.9 	& 1576.0 	&  1605.0 \\
			
			\hline
			
		\end{tabular}
	\end{small}
	\label{table1}
\end{sidewaystable} 
\FloatBarrier

\begin{sidewaysfigure}[h!]
	\centering
	\begin{tabular}{c c c}
		\includegraphics[scale=0.33]{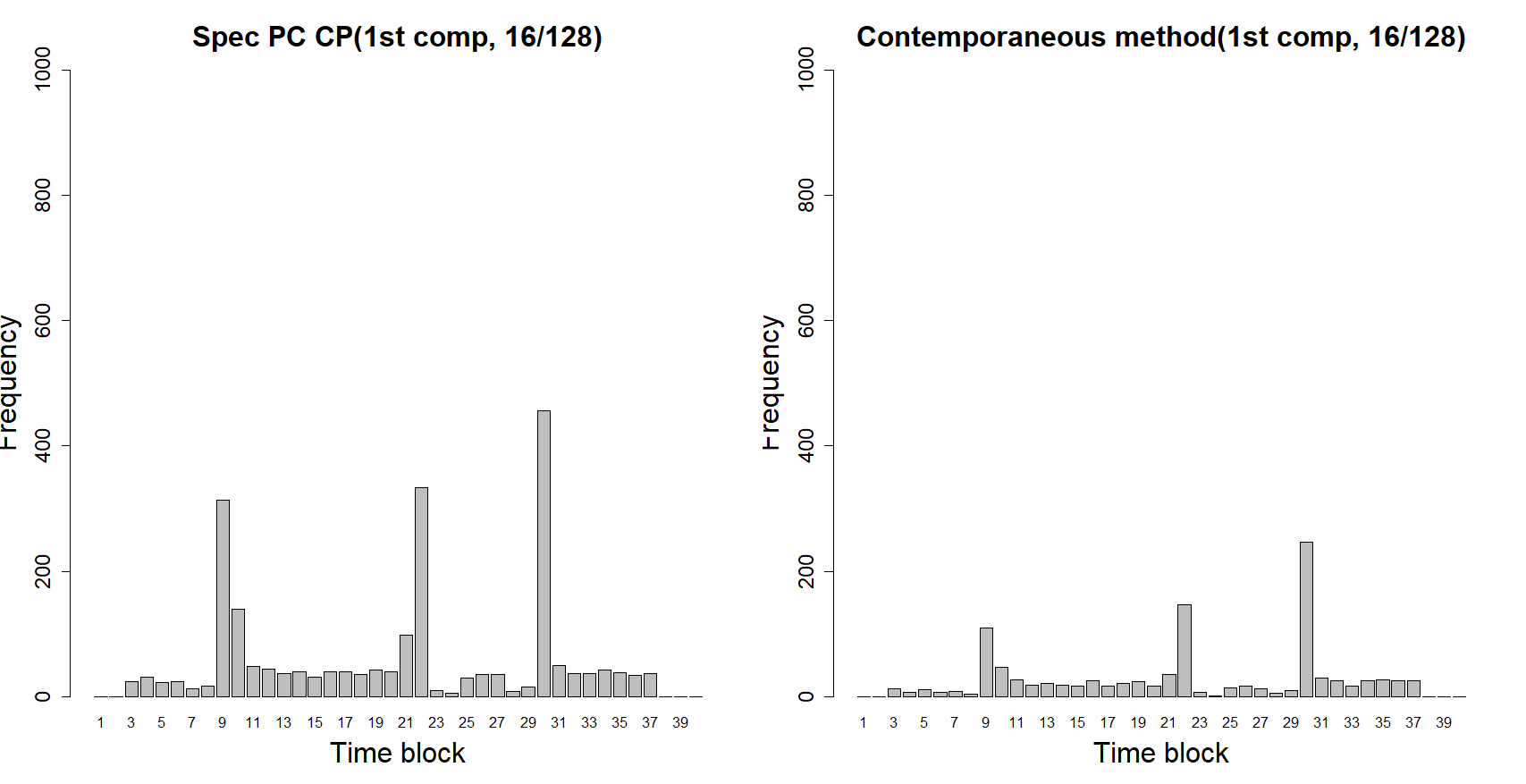}&\includegraphics[scale=0.33]{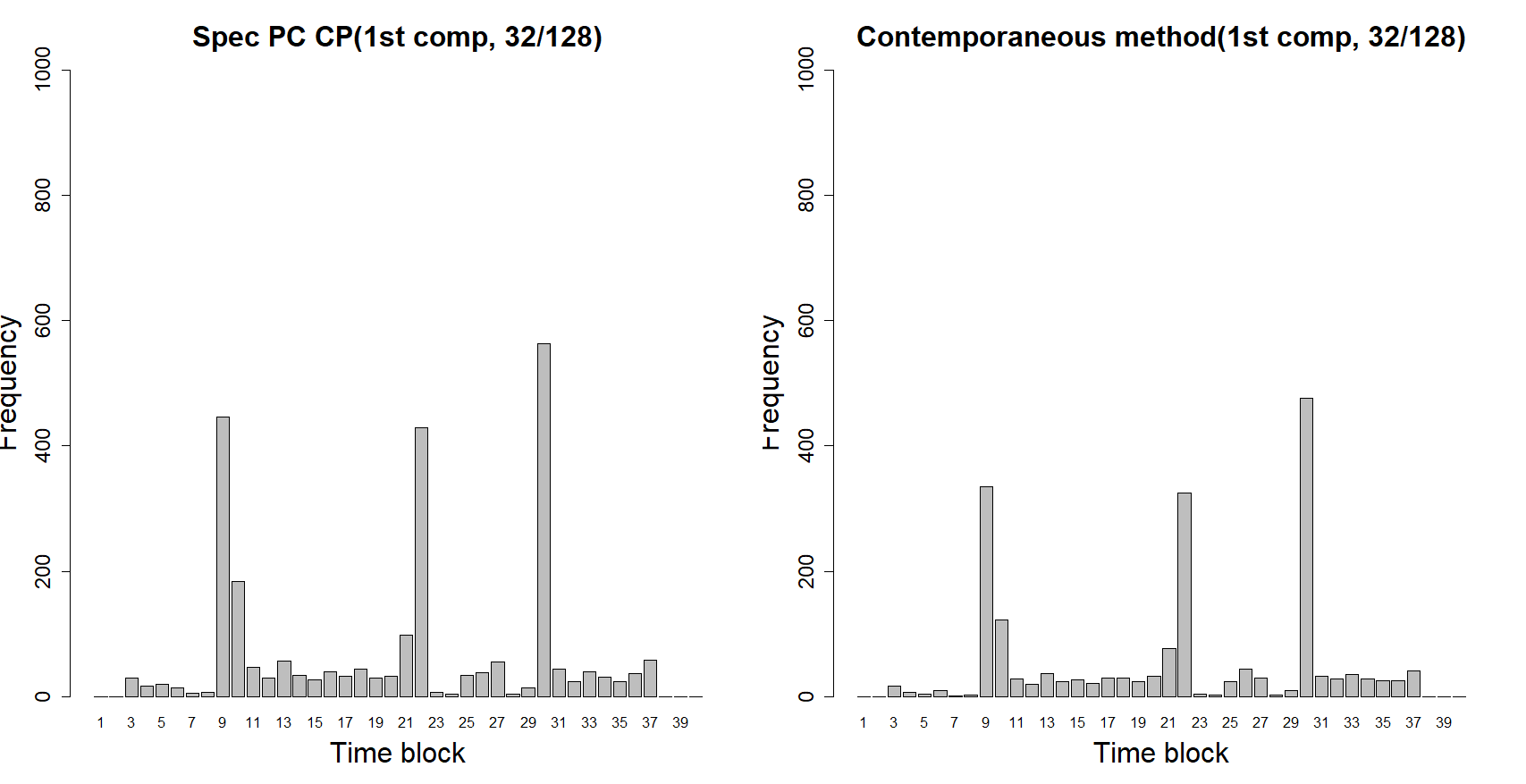}&\includegraphics[scale=0.33]{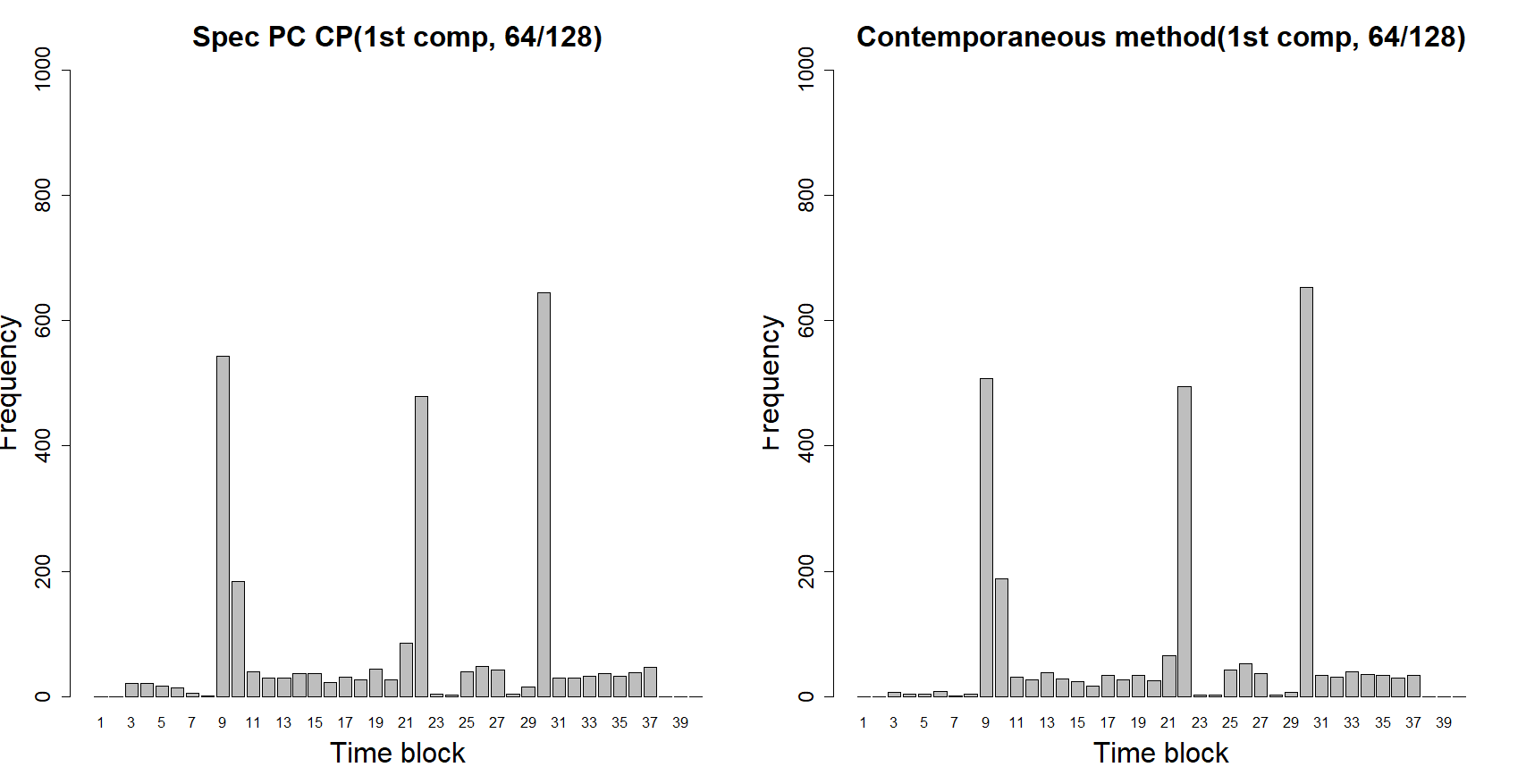}\\
		\includegraphics[scale=0.33]{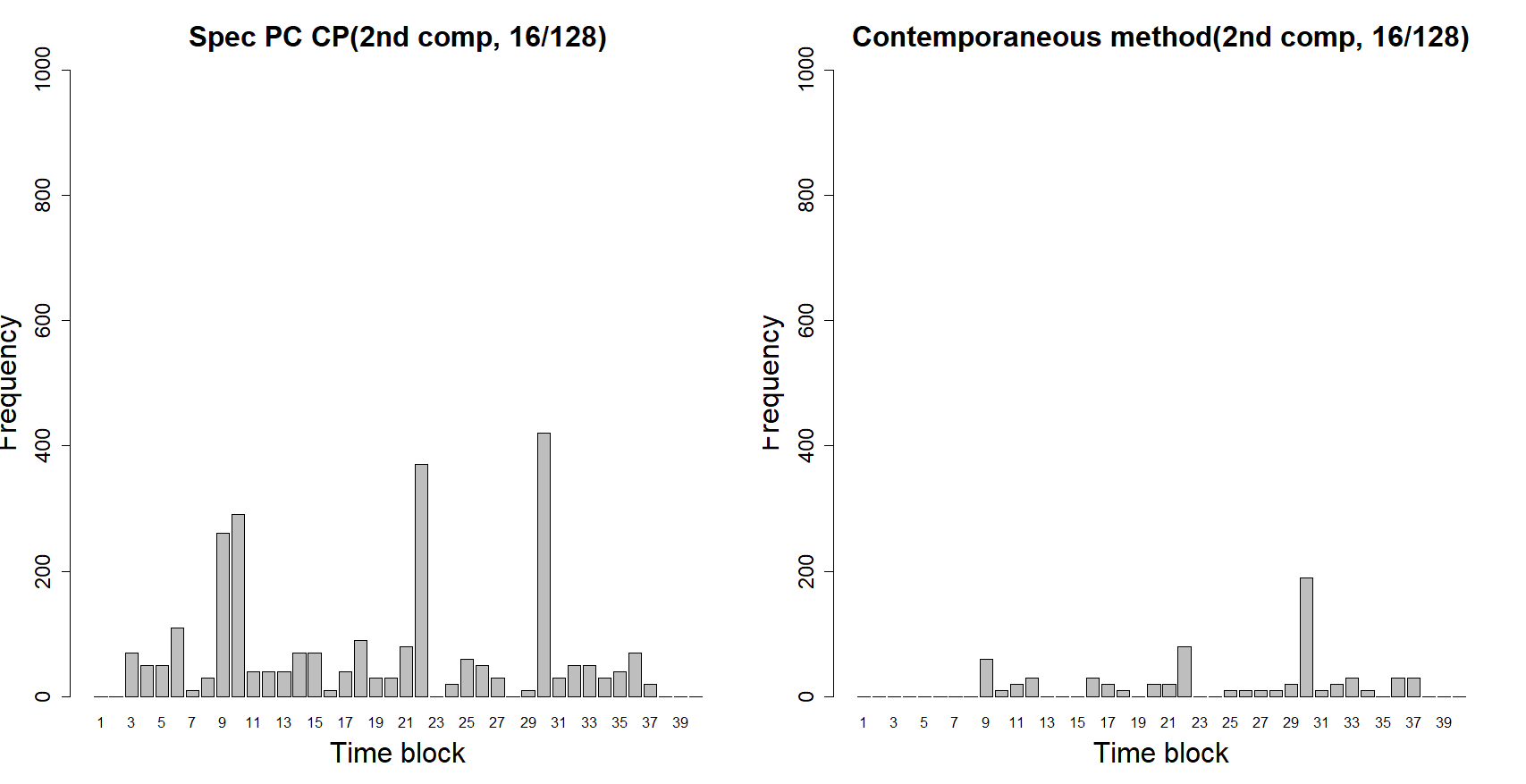}&\includegraphics[scale=0.33]{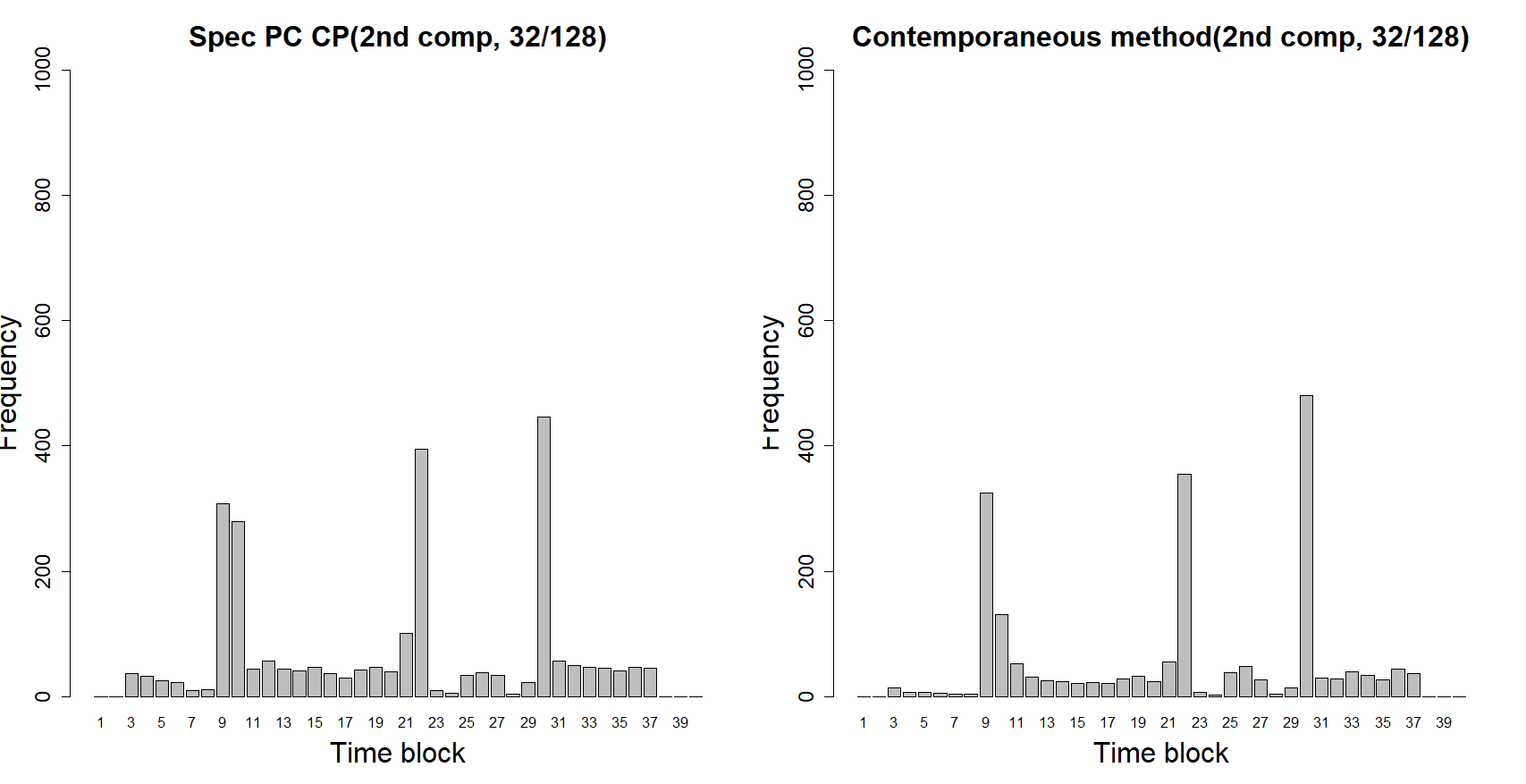}&\includegraphics[scale=0.33]{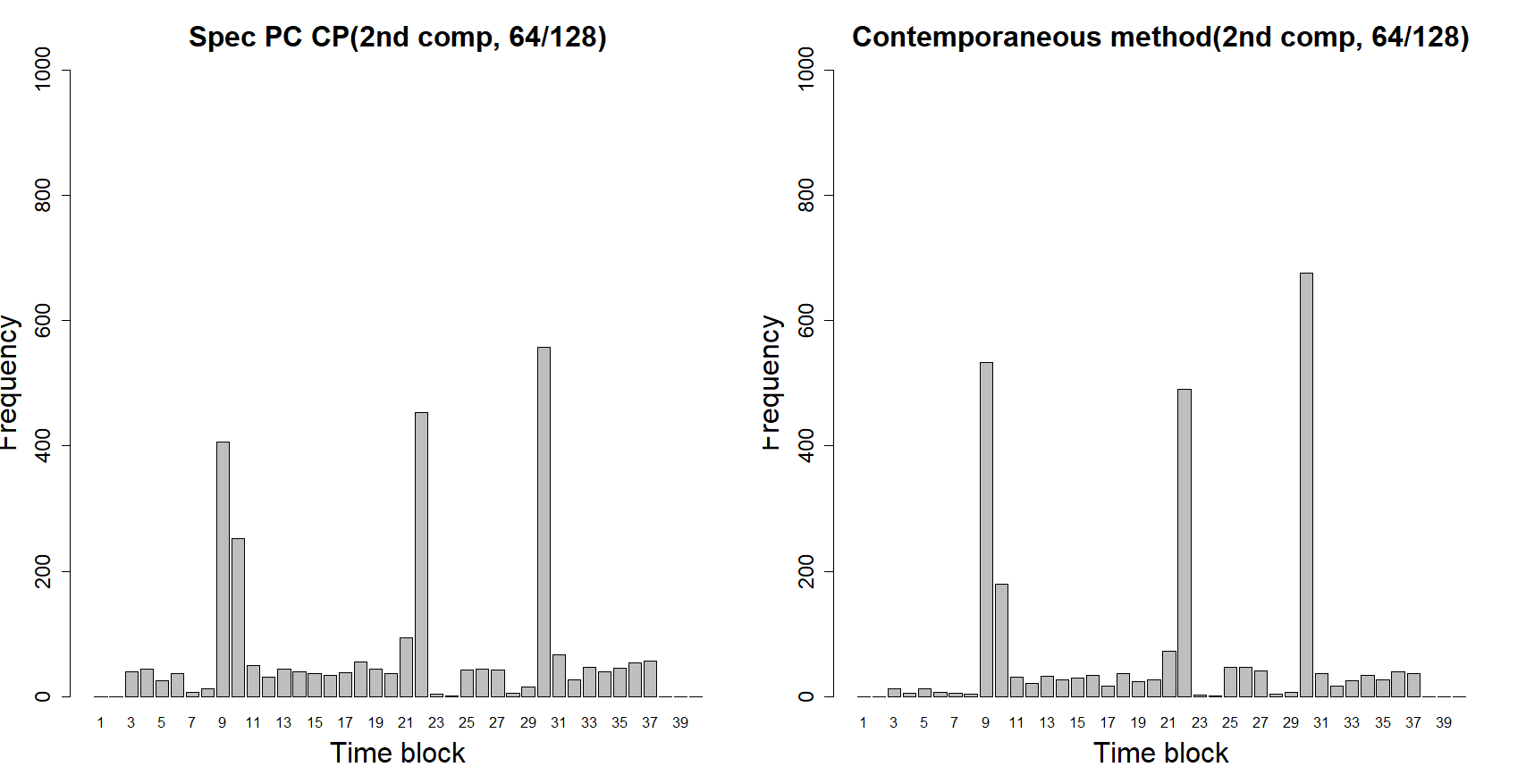}\\
		\includegraphics[scale=0.33]{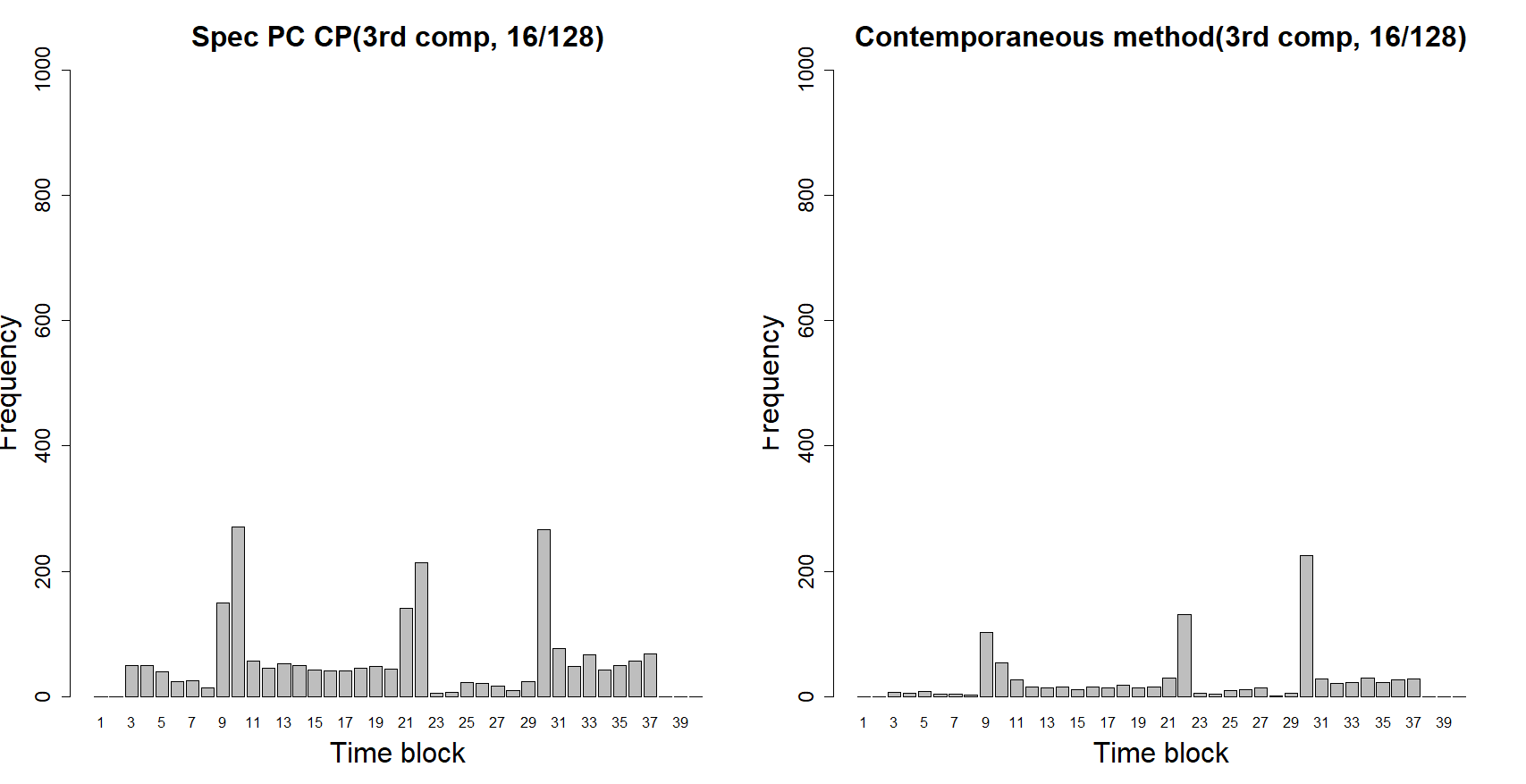}&\includegraphics[scale=0.33]{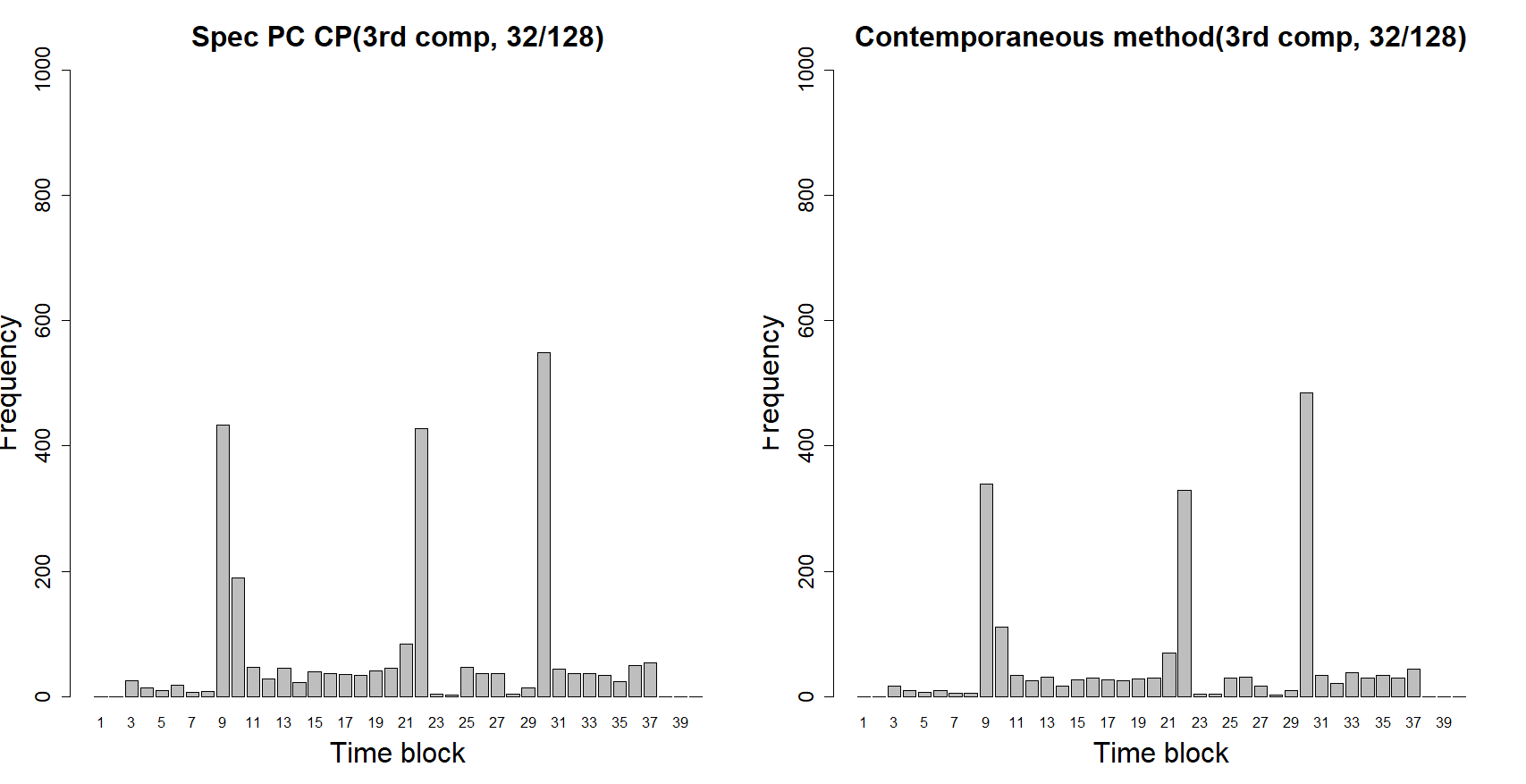}&\includegraphics[scale=0.33]{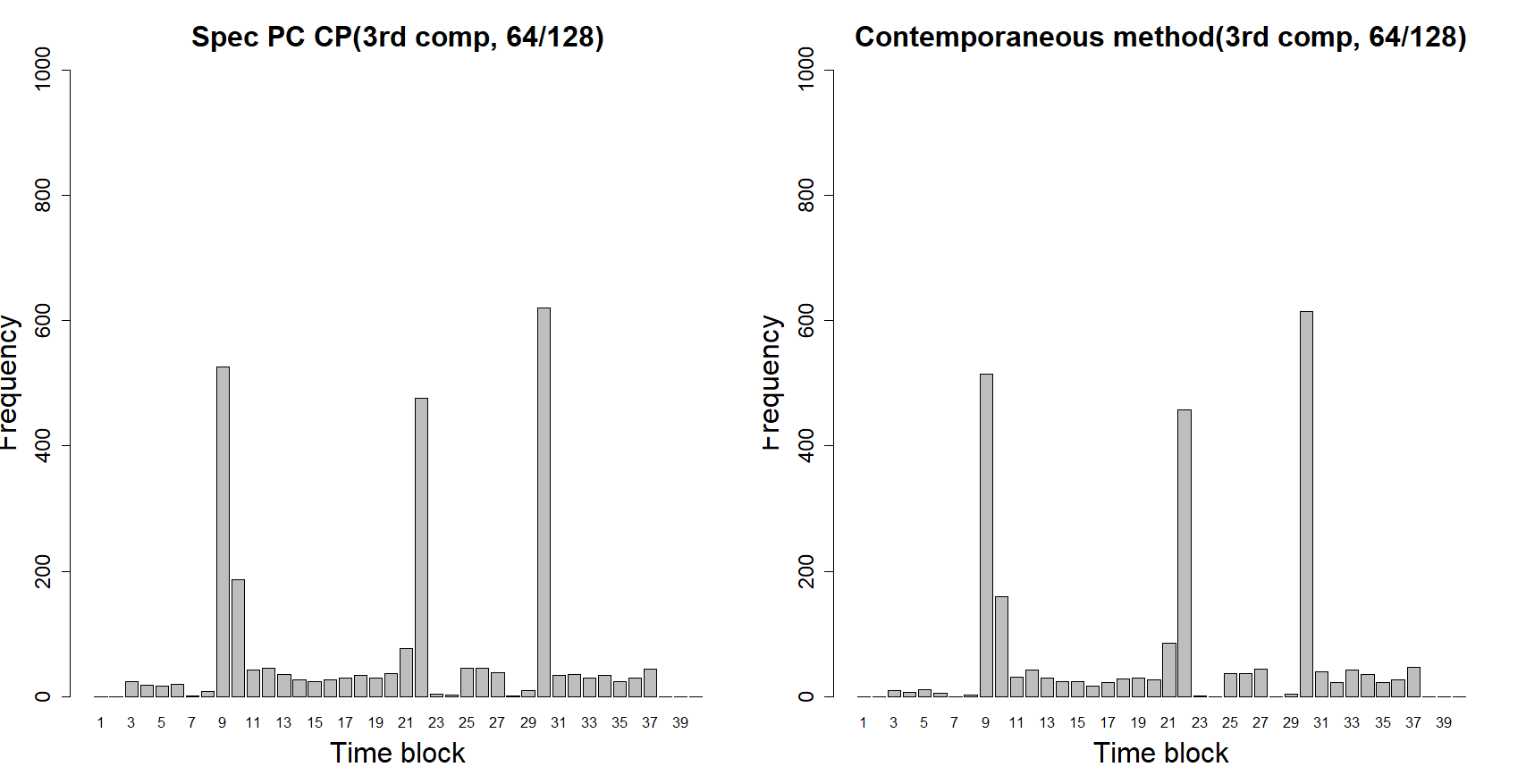}\\
	\end{tabular}
	\caption{Proportion of being detected as change point(out of 1,000). From left to right: 16,32 and 64 channels with change point, respectively. From top to bottom: using the first, second, or third spectral PC.  }
\end{sidewaysfigure}
\FloatBarrier

\begin{sidewaystable}
	\centering
	\caption{Summary of simulation results for the mixed source simulation method. Time series length is $T=1,000$. There is one change point at $t=550$.
		\\ Presented are the detection rates and the mean absolute distances (MADs) between the estimated and true change point.}
	\begin{small}
		\begin{tabular}{llccccccc}
			\hline
			&& \multicolumn{3}{c}{Detection Rate} && \multicolumn{3}{c}{MAD}\\
			\cline{3-5} \cline{7-9}
			\multicolumn{2}{c}{} & 16/128   & 32/128 & 64/128 && 16/128 & 32/128 & 64/128 \\ 
			\hline
			
			1st component	& Spectral PCA 	&	\textbf{0.66}	&  {0.82}   &{0.90}  	&&	63.6        & 52.0	&  54.0 \\
			& Contemporaneous Method   	&      0.49        &  0.80   & 0.92   	&&    70.9 	& 54.2 	&  105.4 \\

			2nd component	& Spectral PCA 	&	\textbf{0.83}	&  {0.88}   &{0.91}  	&&	57.8        & 56.6 	&  56.7 \\
			& Contemporaneous Method   	&      0.52        &  0.79   & 0.92   	&&    76.9 	& 58.1 	&  55.1 \\

			3rd component	& Spectral PCA 	&	\textbf{0.68}	&  {0.72}   &{0.91}  	&&	61.8        & 61.0	&  55.6 \\
			& Contemporaneous Method   	&      0.52        &  0.80   & 0.93   	&&    76.9 	& 57.5 	&  54.2 \\
			\hline
			& Structual Break   	&      0.69        & 0.63   & 0.23 	&&     439.7 & 439.7 	&  439.1 \\
			\hline
			& Sparsified Binary Segmentation   	&      0.24        & 0.28   & 0.36 	&&     10.2 & 18.0 	&  21.5 \\
			
			\hline
		\end{tabular}
		
	\end{small}
	\label{table1}
\end{sidewaystable} 
\FloatBarrier

\begin{sidewaysfigure}[h!]
	\centering
	\begin{tabular}{c c c}
		\includegraphics[scale=0.33]{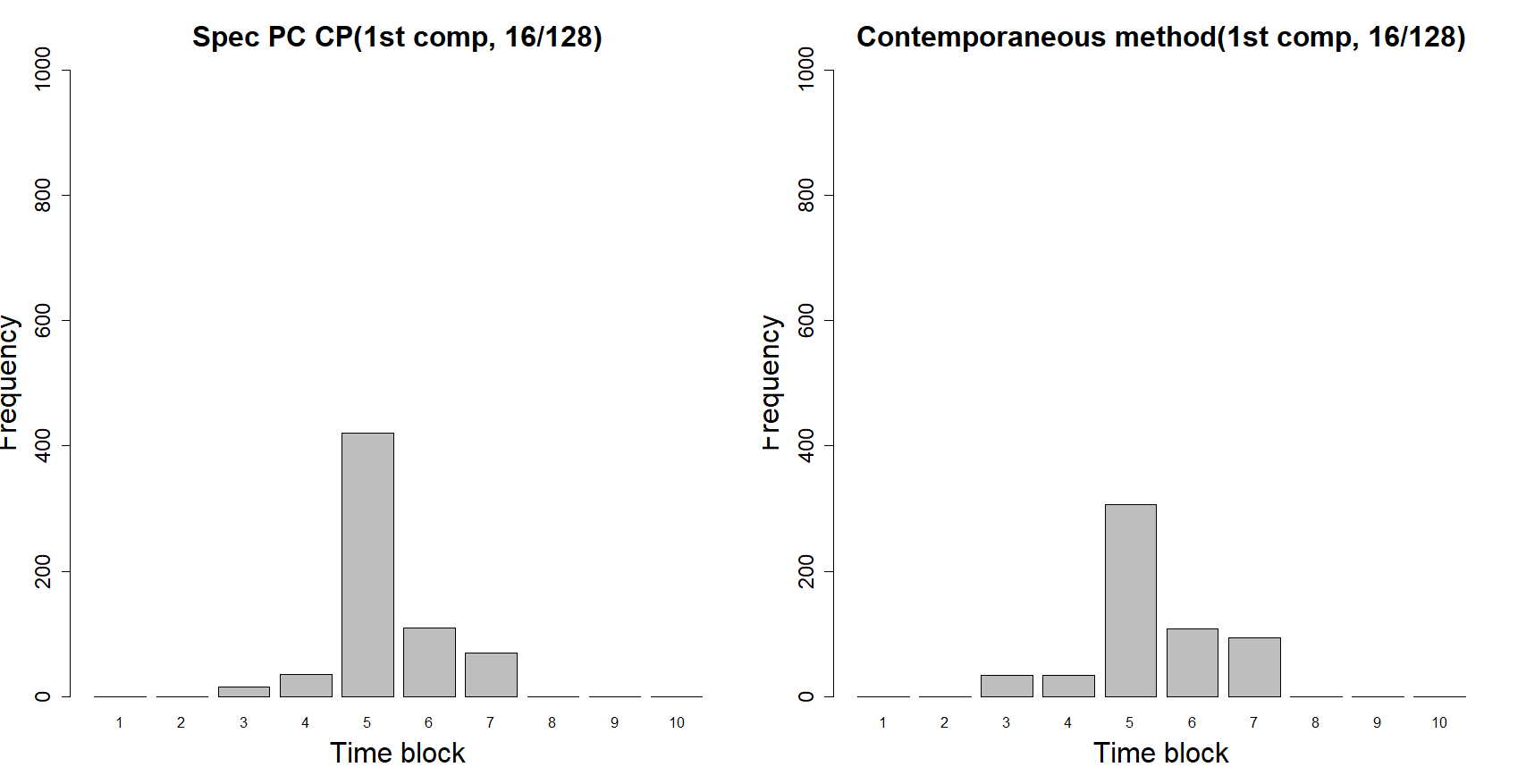}&\includegraphics[scale=0.33]{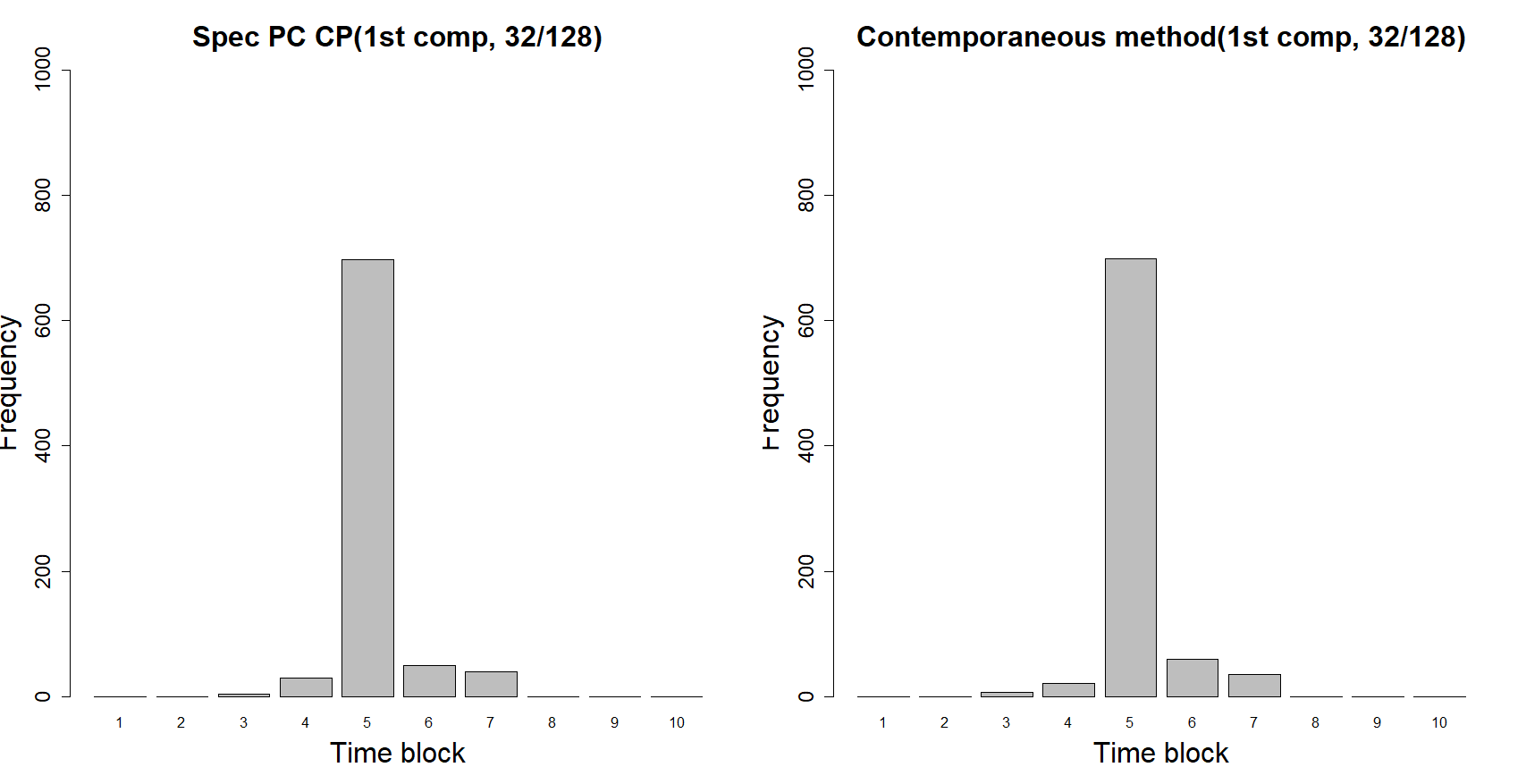}&\includegraphics[scale=0.33]{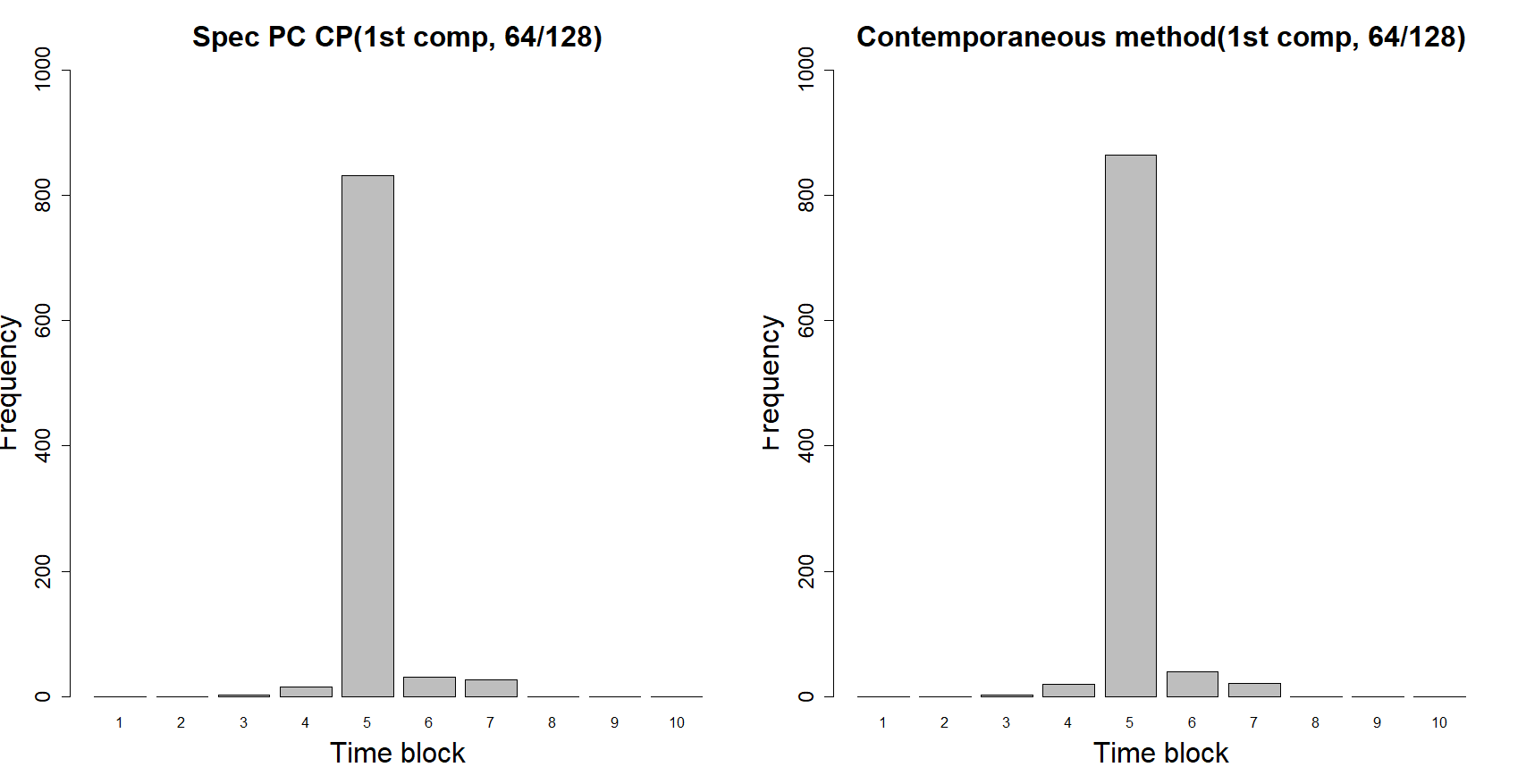}\\
		\includegraphics[scale=0.33]{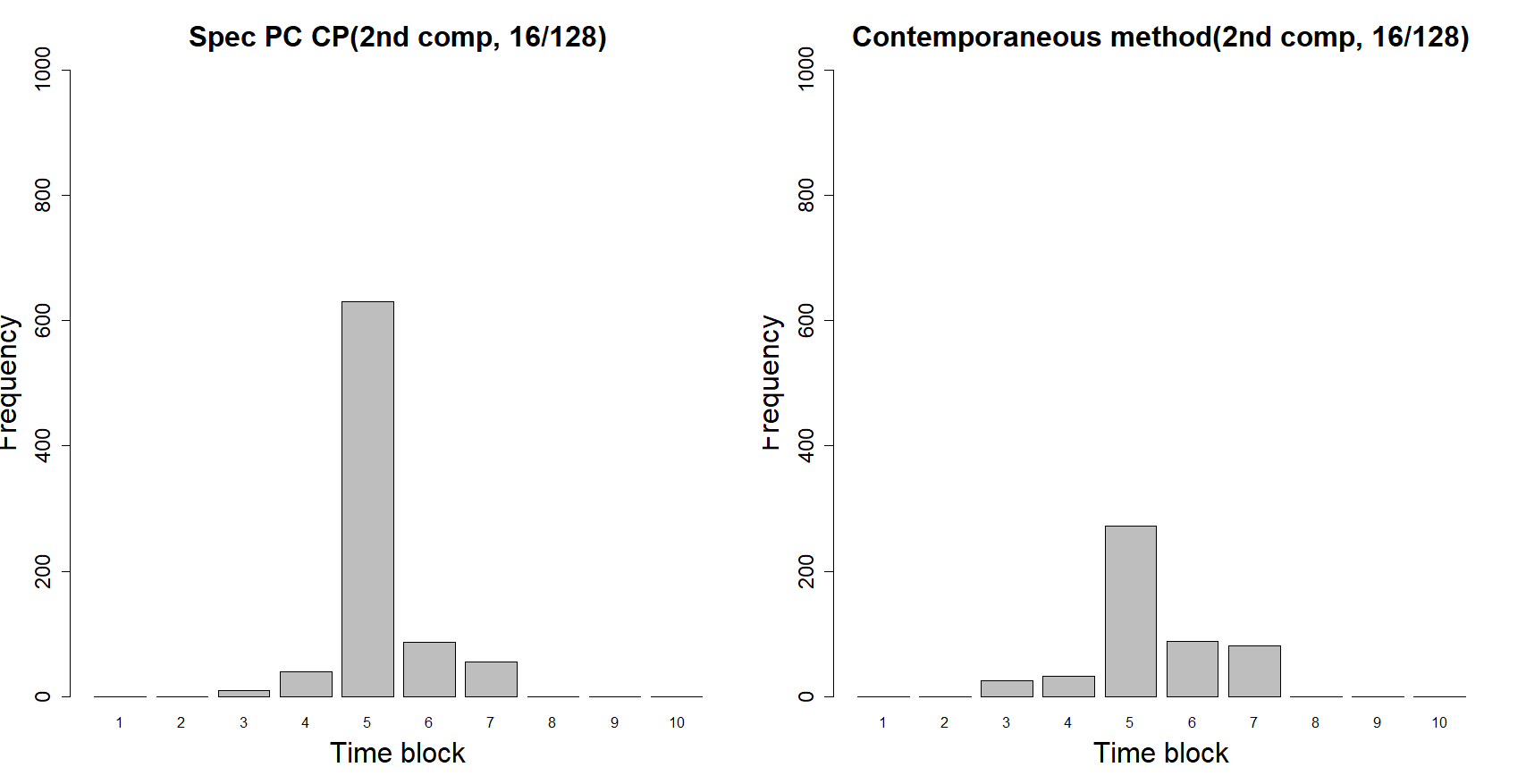}&\includegraphics[scale=0.33]{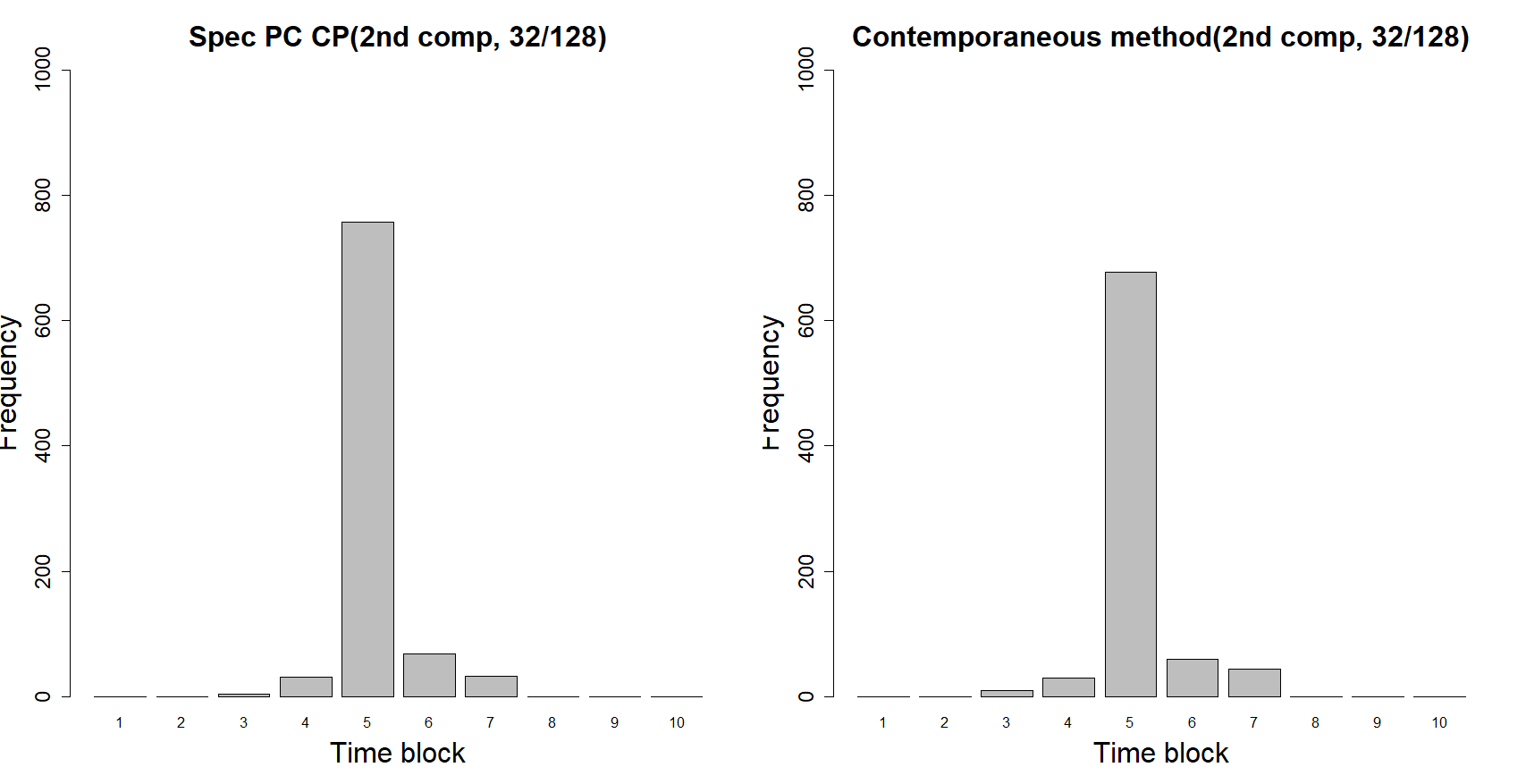}&\includegraphics[scale=0.33]{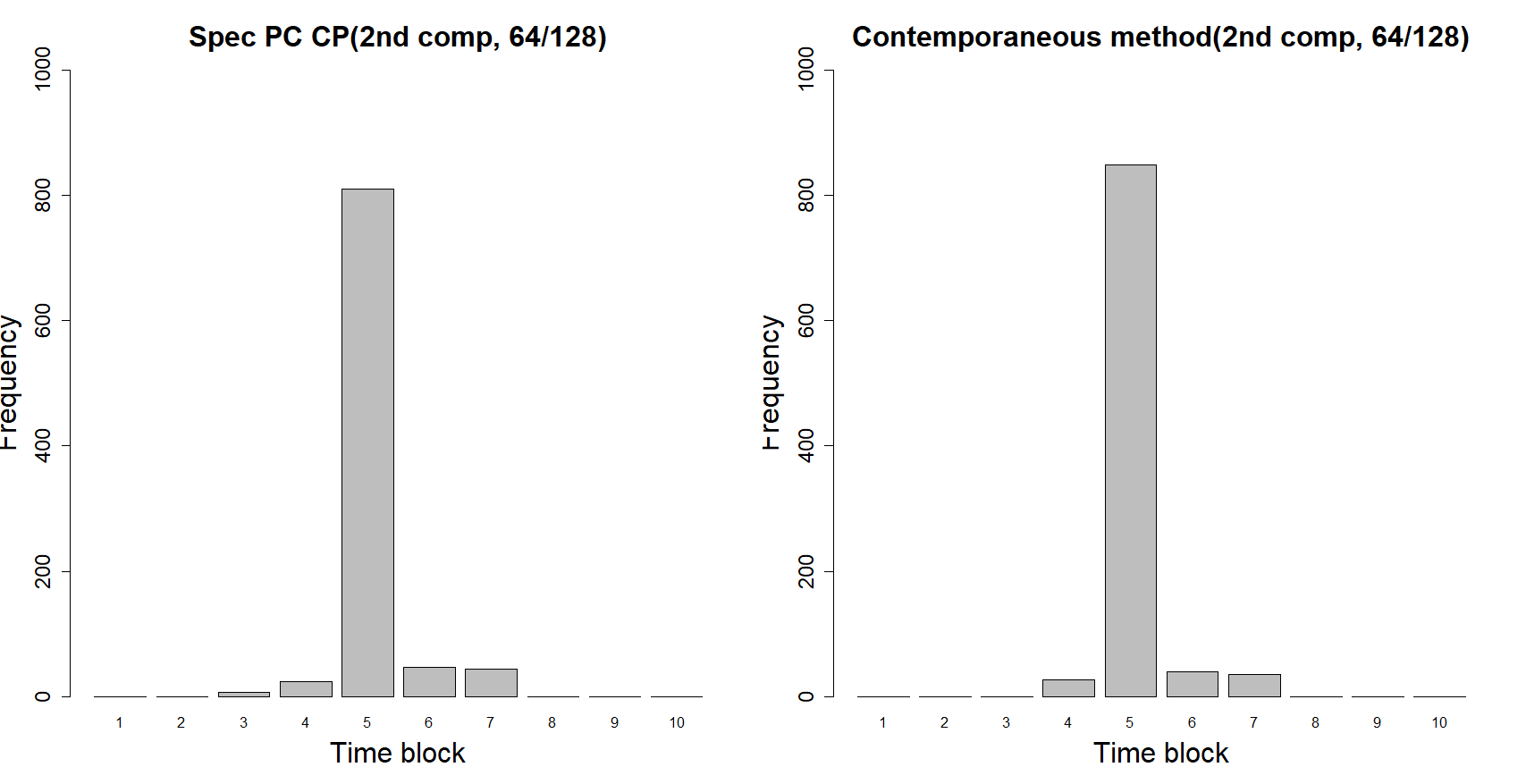}\\
		\includegraphics[scale=0.33]{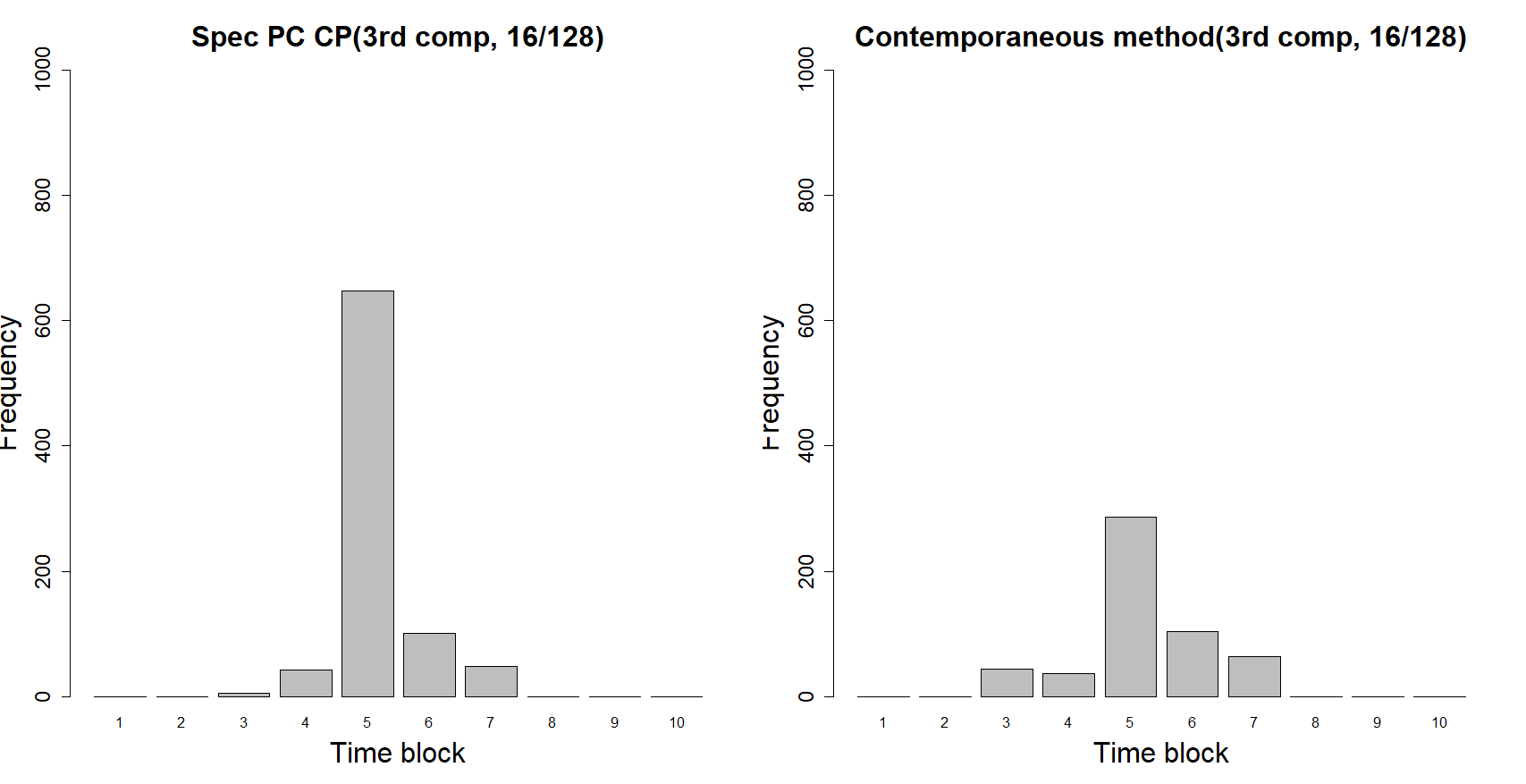}&\includegraphics[scale=0.33]{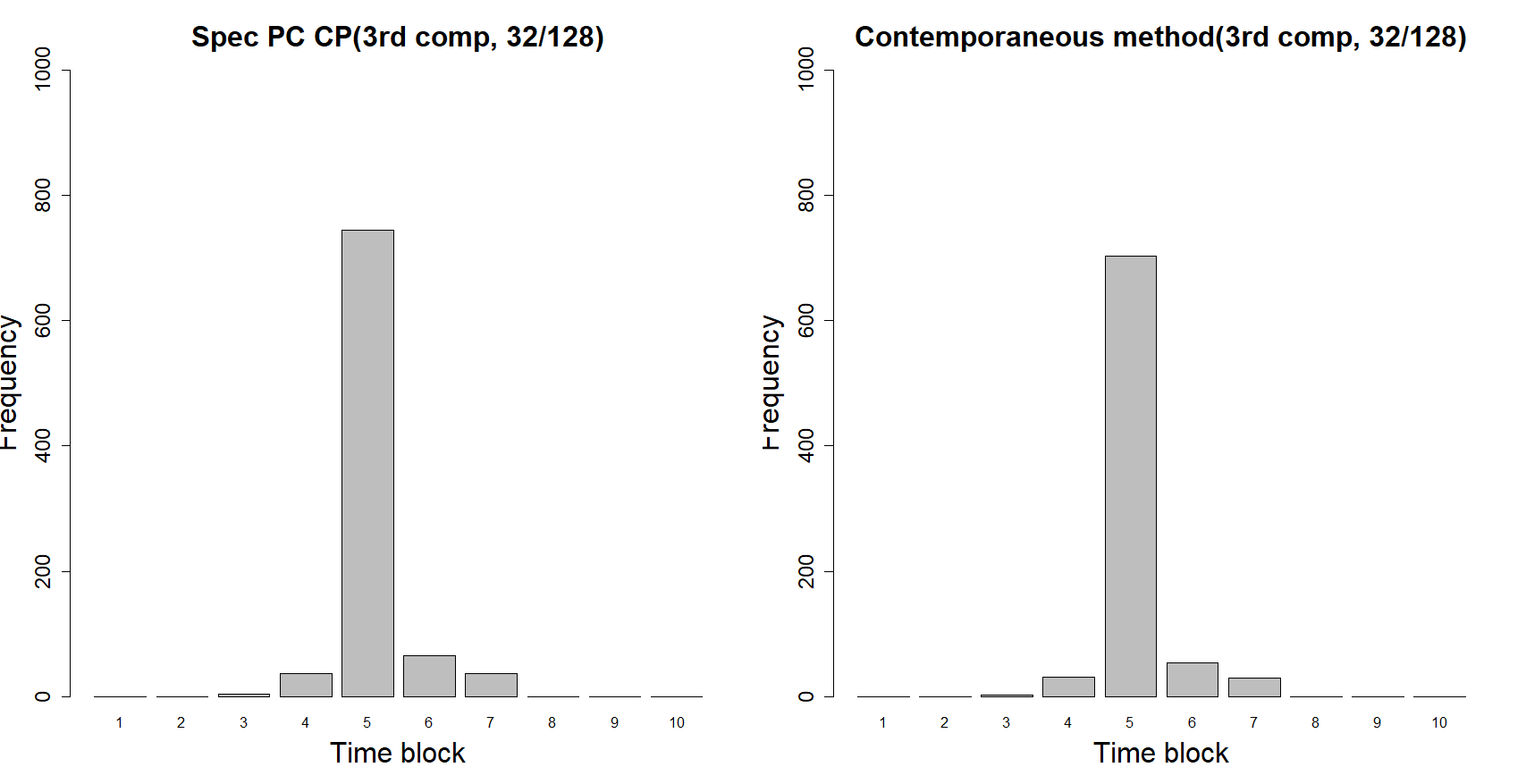}&\includegraphics[scale=0.33]{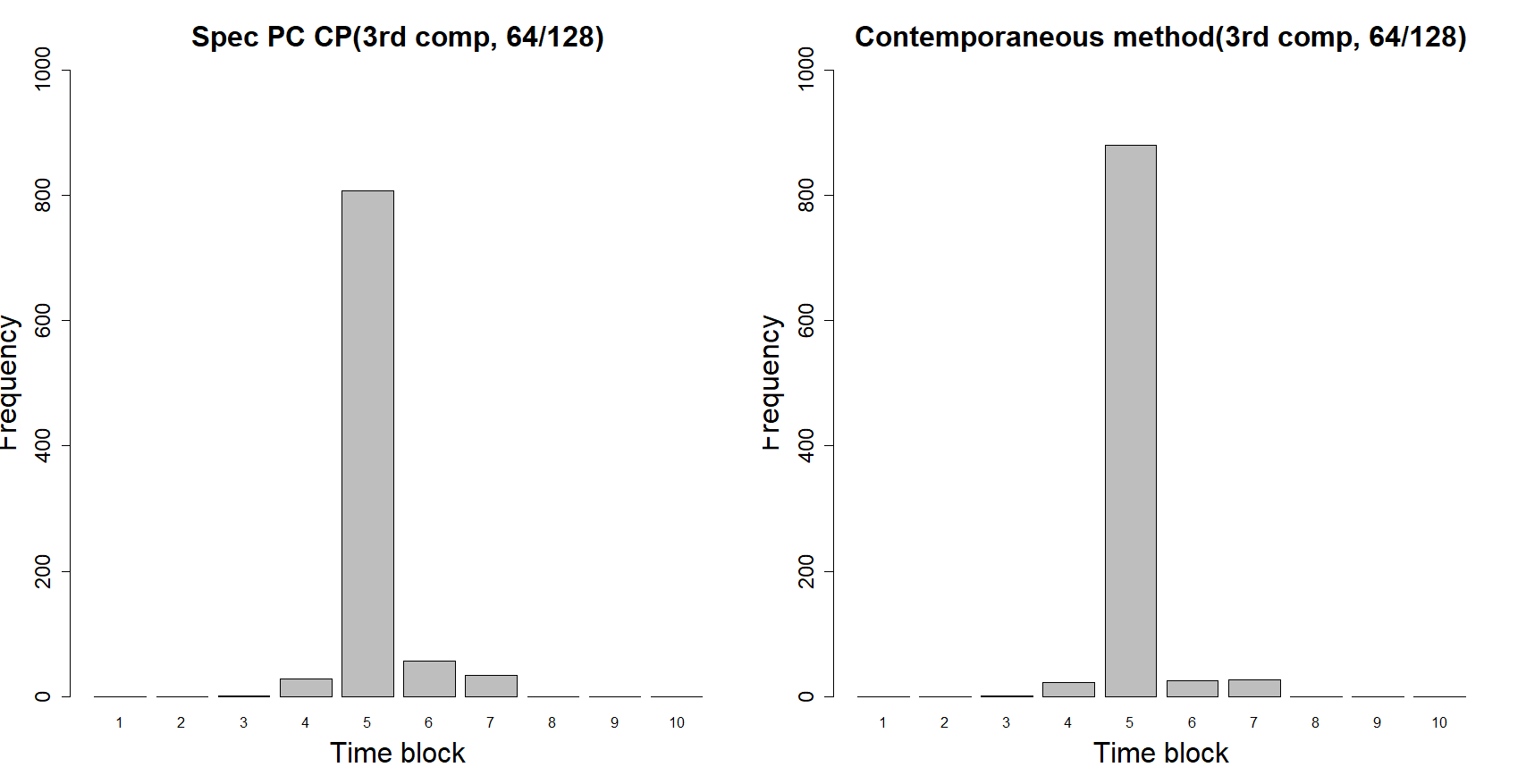}\\
	\end{tabular}
	\caption{Proportion of being detected as change point(out of 1,000). From left to right: 16,32 and 64 channels with change point, respectively. From top to bottom: using the first, second, or third spectral PC.  }
\end{sidewaysfigure}
\FloatBarrier	

	\section{Data Analysis}
\subsection*{EEG Data}
We applied our two-stage change point detection method to two data sets. The first is a seizure recording which captured brain activity of a subject monitored at the epilepsy center at the University of Michigan. The EEG data was sampled at 100Hz and lasted for about 500 seconds with a total length of 50,000 (Sch\"oder and Ombao, 2019). The data was recorded at 31 channels. The placement of the scalp electrodes is illustrated in Figure 7. The abbreviation correspond to different locations on the scalp. For example, Fp means frontal polar and C means central.
Figure 8 shows the time series for six channels where the seizure happening time are similar among different channels.
\begin{figure}[h!]
	\begin{center}
		\includegraphics[scale=1.2]{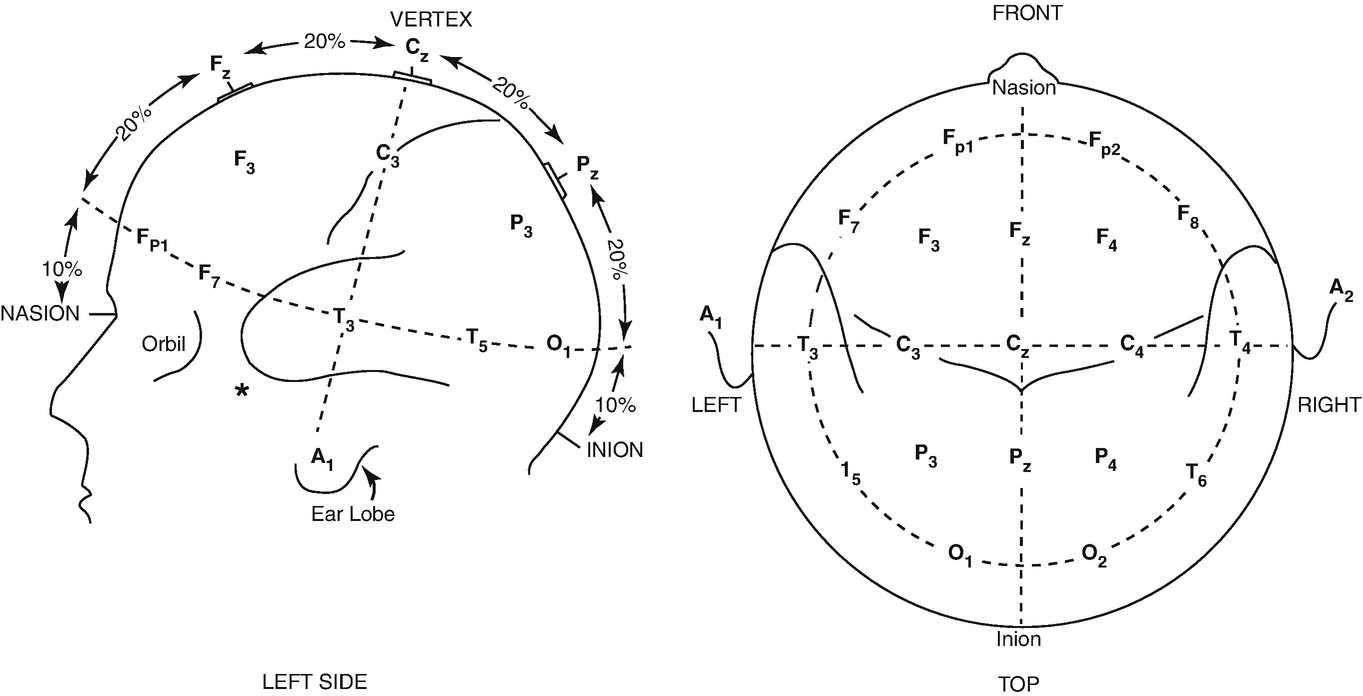}
	\end{center}
	\caption{EEG scalp topography}
\end{figure}
\begin{figure}[h!]
	\begin{center}
		\includegraphics[scale=1.4]{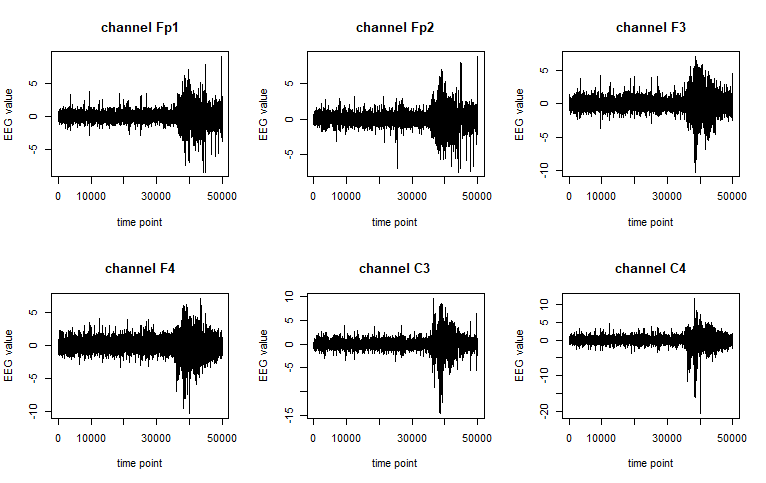}
	\end{center}
	\caption{Six channels of the EEG seizure data (channel Fp1, Fp2, F3, F4, C3, C4)}
\end{figure}
\FloatBarrier

We first examine the time-frequency plots of the first three spectral PCs (Figure 9) where
darker color indicates higher weights at certain frequency and time points. It has been known that the low-frequency energy is more varying during epileptic seizure (Sch\"oder and Ombao, 2019). Therefore, it is of interest to use delta band to detect change points. 

Figure 10 shows the detected change points using the first three spectral PCs. The red dotted lines denote the location of estimated change points in delta band. We have detected a few pre-ictal change points in all three components at the beginning of the recording. The first component captures multiple change points while the second and third component each detects one change point before seizure took place. The result suggests that it might be possible to build early warning systems on seizure. 
The other group of change points is identified right before and during seizure onset. All of the three components capture several changes around $t=40,000$, which agree with visual inspection on the frequency-time plot and the findings of neurologists. This demonstrates that our method can detect epileptic seizure well on this data.

To understand how many spectral PC should be analyzed, we plotted the variance proportion plot (Figure 10), which shows that the first two components explain over 90\% of the total variance.	Thus, analyzing the first three principal component is adequate for this particular dataset.                

We also conduct analysis using the other comparison methods. Figure 11 shows the results using Contemperaneous Method and it did not detect the pre-ictal change points using the first two components. In figure 12 it can be seen that the Structural Break Method can only detect a change point at the end of the series and the Sparsified Binary Segmentation cannot detect the pre-ictal change point either.

\begin{figure}[h!]
	\centering
	\begin{tabular}{c c }
		\includegraphics[scale=0.5]{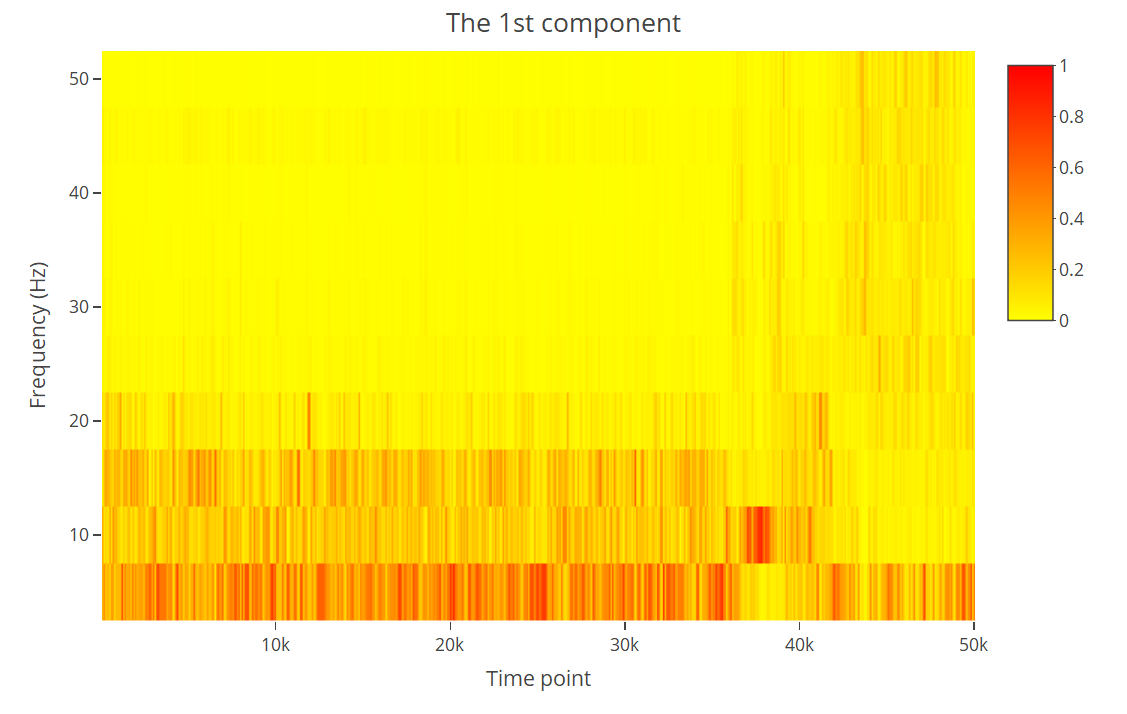}&\includegraphics[scale=0.5]{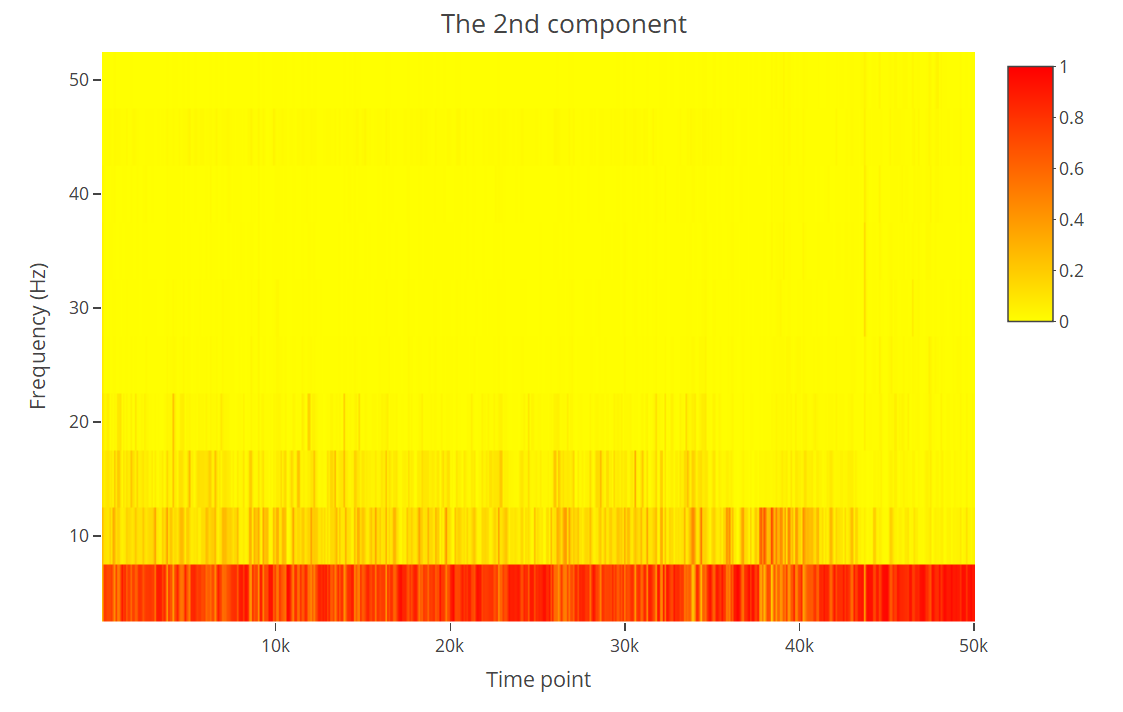}\\
		\includegraphics[scale=0.5]{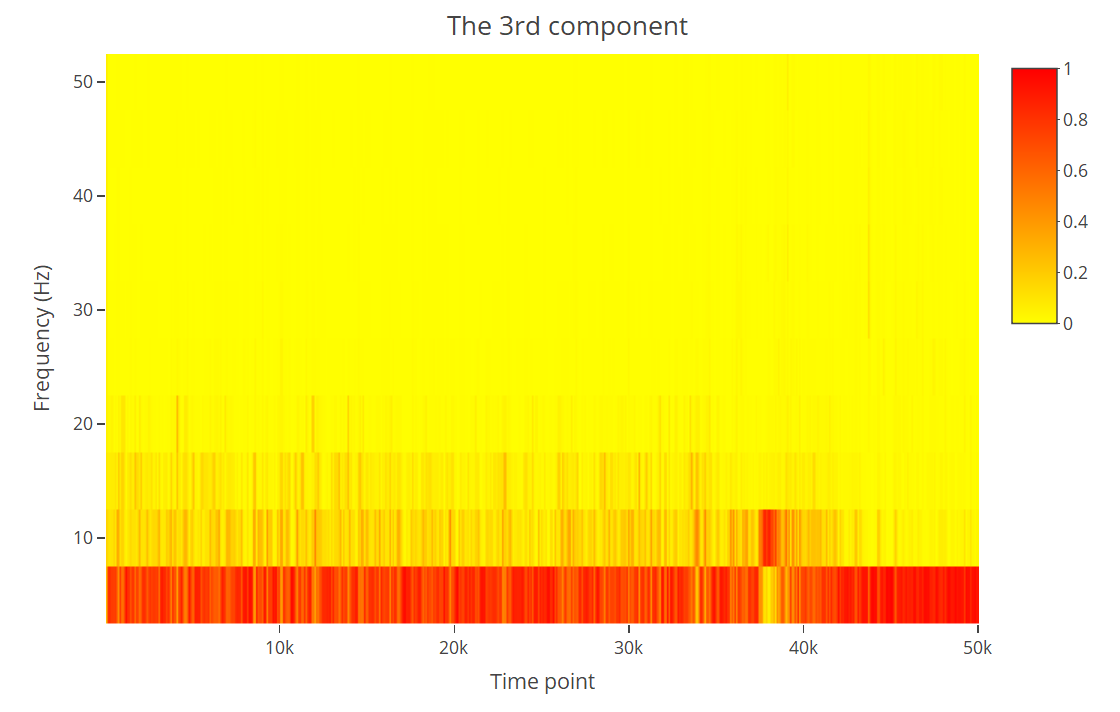}
	\end{tabular}
	\caption{The time-frequency plot of the first three spectral PCs. The x-axis denotes the time; the y-axis denoted the frequency from 0 to 50 Hertz. }
\end{figure}

\begin{figure}[h!]
	\centering
	\begin{tabular}{c c }
		\includegraphics[scale=0.7]{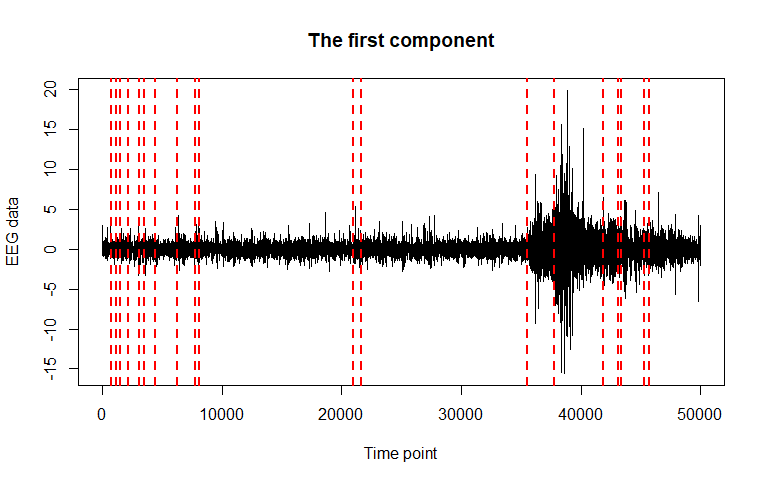}&\includegraphics[scale=0.7]{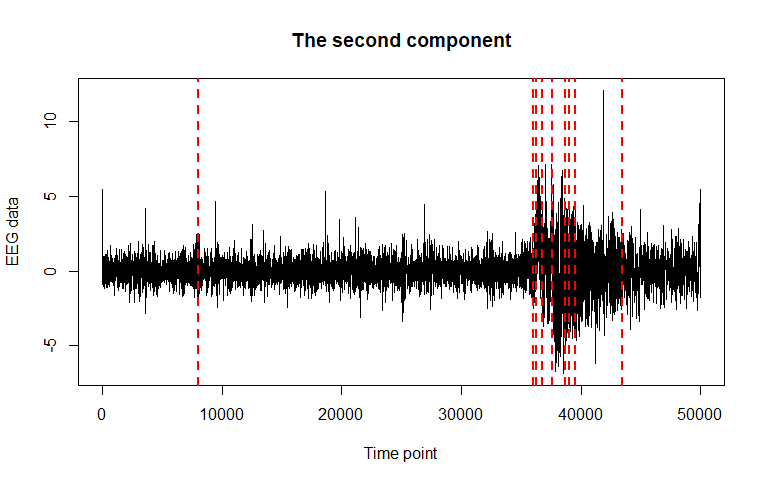}\\
		\includegraphics[scale=0.7]{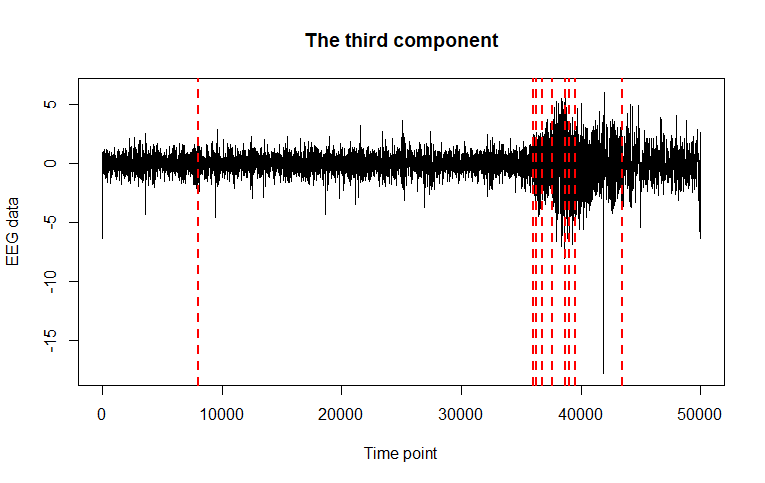}&\includegraphics[scale=0.7]{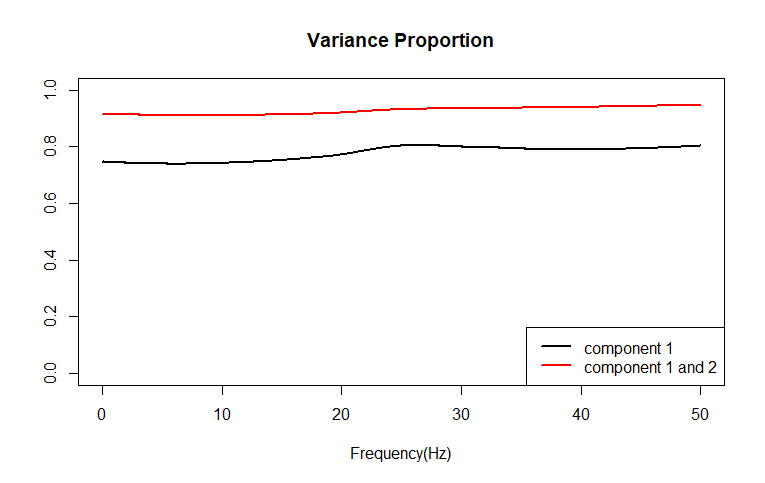}
	\end{tabular}
	\caption{The estimated change point using the first, second or third spectral PC. The red dotted lines denote the locations of estimated change points. Bottom right: The variance explained by the first component only and first and second components together.}
\end{figure}
\FloatBarrier

\begin{figure}[h!]
	\centering
	\begin{tabular}{c c }
		\includegraphics[scale=0.7]{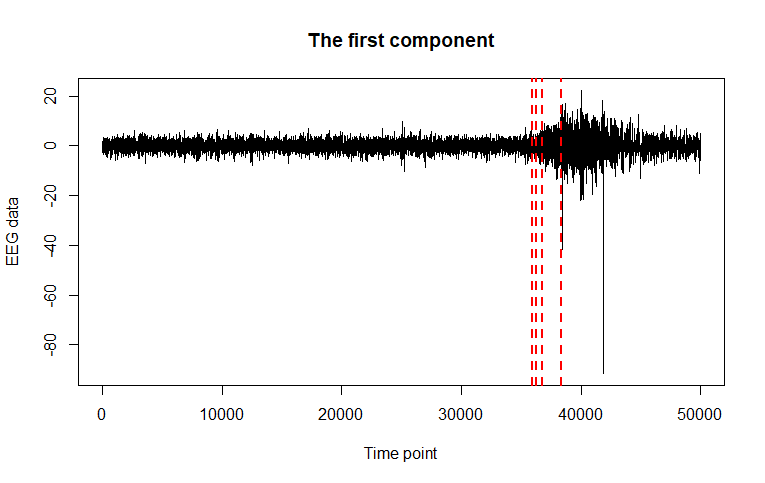}&\includegraphics[scale=0.7]{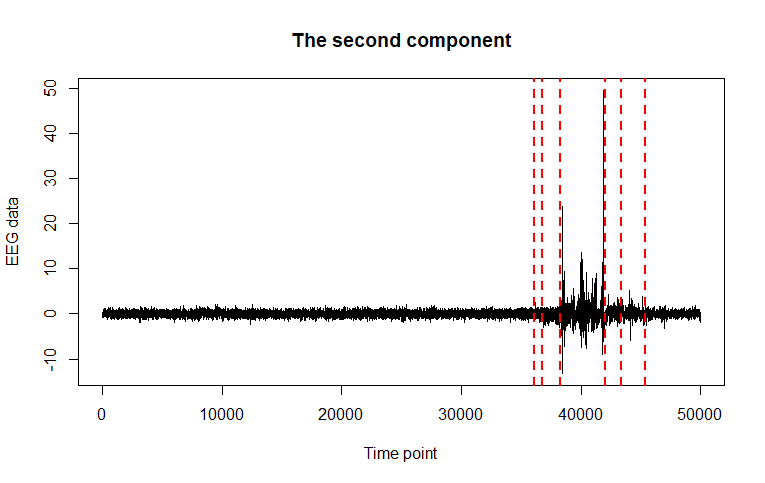}\\
		\includegraphics[scale=0.7]{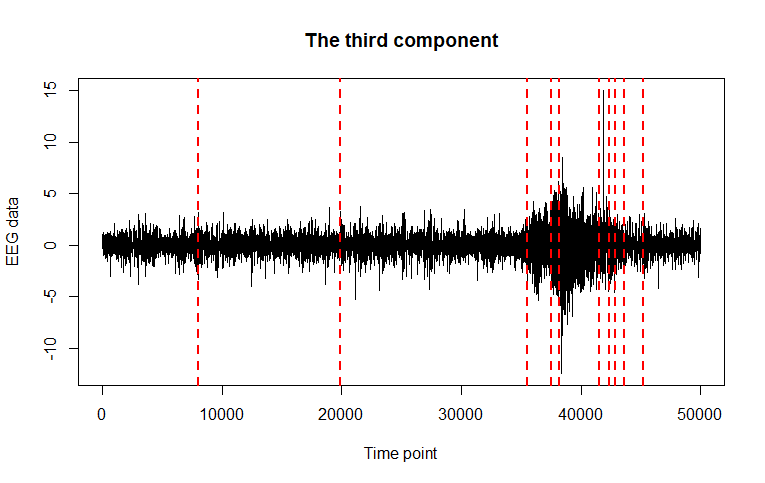}&
	\end{tabular}
	\caption{The estimated change point using the first, second or third component of Contemperaneous Method. The red dotted lines denote the location of estimated change points.}
\end{figure}
\FloatBarrier

\begin{figure}[h!]
	\centering
	\begin{tabular}{c c }
		\includegraphics[scale=0.7]{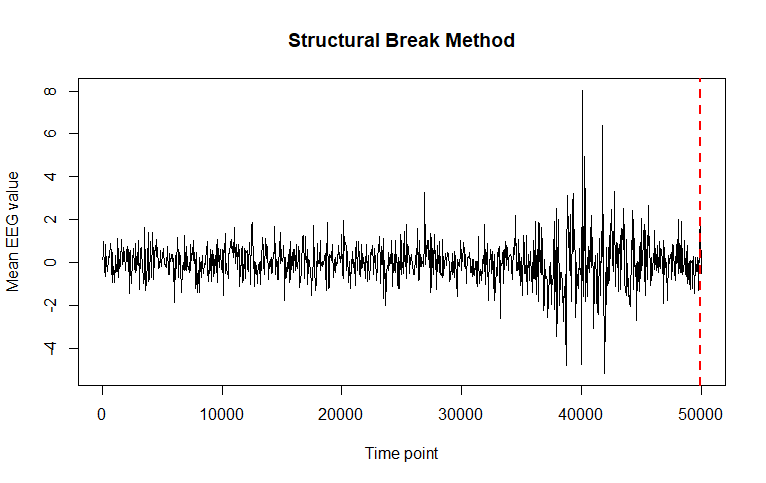}&\includegraphics[scale=0.7]{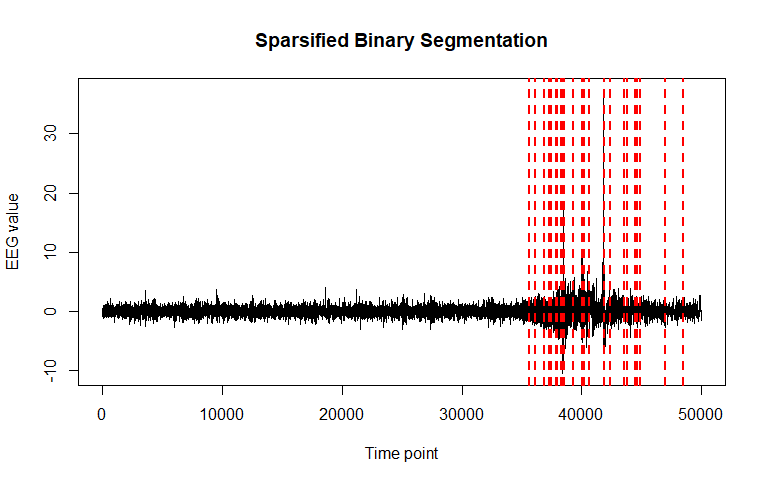}
	\end{tabular}
	\caption{The mean EEG values across all the channels and estimated change point using Structural Break Method and Sparsified Binary Segmentation (Downsampling is performed by keeping only every 50th sample to apply the Structural Break Method). The red dotted lines denote the location of estimated change points. }
\end{figure}
\FloatBarrier

\subsection*{Stock Data} 

We also applied our Spec PC-CP method to 
the log returns of the daily closing values of 9 representative $S\&P$ 100 stocks which has been analyzed in previous work (Barigozzi, Cho and Fryzlewicz, 2014). The stock data was observed between 4 January 2000 and 10 August 2016 from Yahoo finance. The total length is 4,177 days. We analyzed the following 9 stocks: AXP (American Express), BAC (Bank of America), BK(The Bank of New York Mellon), C (Citigroup), COF (Capital One Financial), GS (Goldman Sachs), JPM (JPMorgan), MS (Morgan Stanley), WFC (Wells Fargo). Figure 13 shows the stock prices of Bank of America and JP Morgan.

\begin{figure}[h!]
	\centering
	\begin{tabular}{c c}
		\includegraphics[scale=0.8]{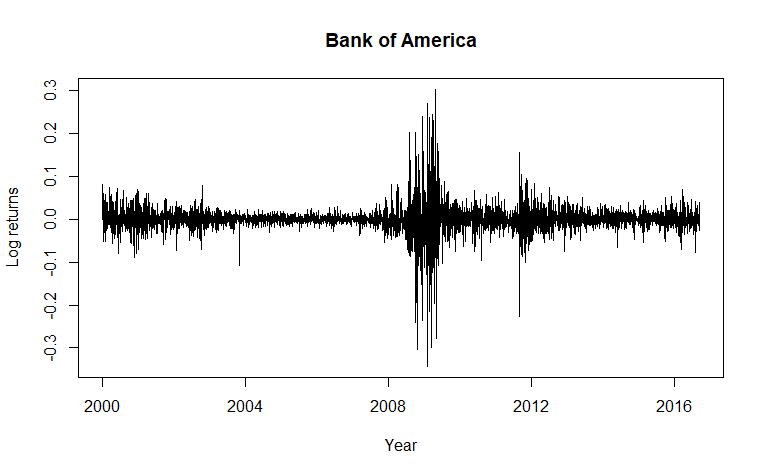}&
		\includegraphics[scale=0.8]{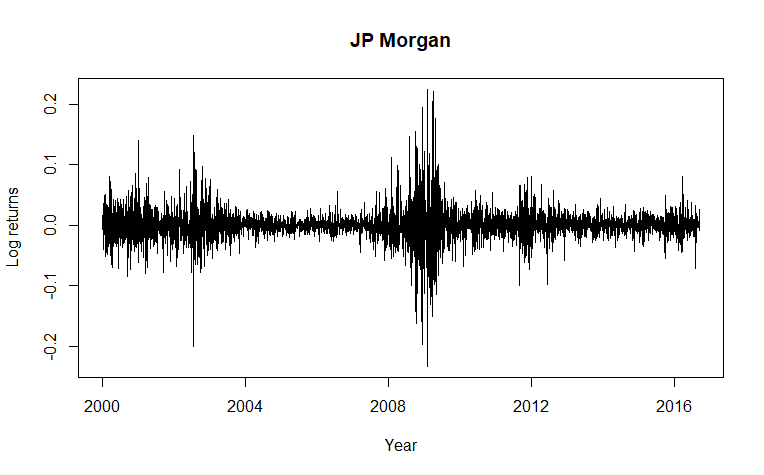}
	\end{tabular}
	\caption{Daily log returns of Bank of America and JPMorgan between 2000 and 2016}
\end{figure}
\FloatBarrier

We used the first two spectral principal components for change point detection. 
Figure 14 gives the frequency-time plots of two stocks and the first two spectral PCs. Changes appear in multiple frequency bands in this case. 
The detected change points are shown in Figure 15.
Most of the change points are around events that might have impact on the financial market. For example, the start of the Iraq War in 2003, 
The financial crisis in 2008, the Greek and EU sovereign debt crisis in 2011 and 2015. Both the first and second spectral principal components are able to detect the change points in a neighbourhood of the above events. These results demonstrates that our method also work well for stock data. Figure 16 shows the results using the comparison methods. The Contemperaneous Method has comparable estimates while the other two methods fail to detect events such as the financial crisis in 2008 and the debt crisis in 2015.


\begin{figure}[h!]
	\centering
	\begin{tabular}{c c}
		\includegraphics[scale=0.37]{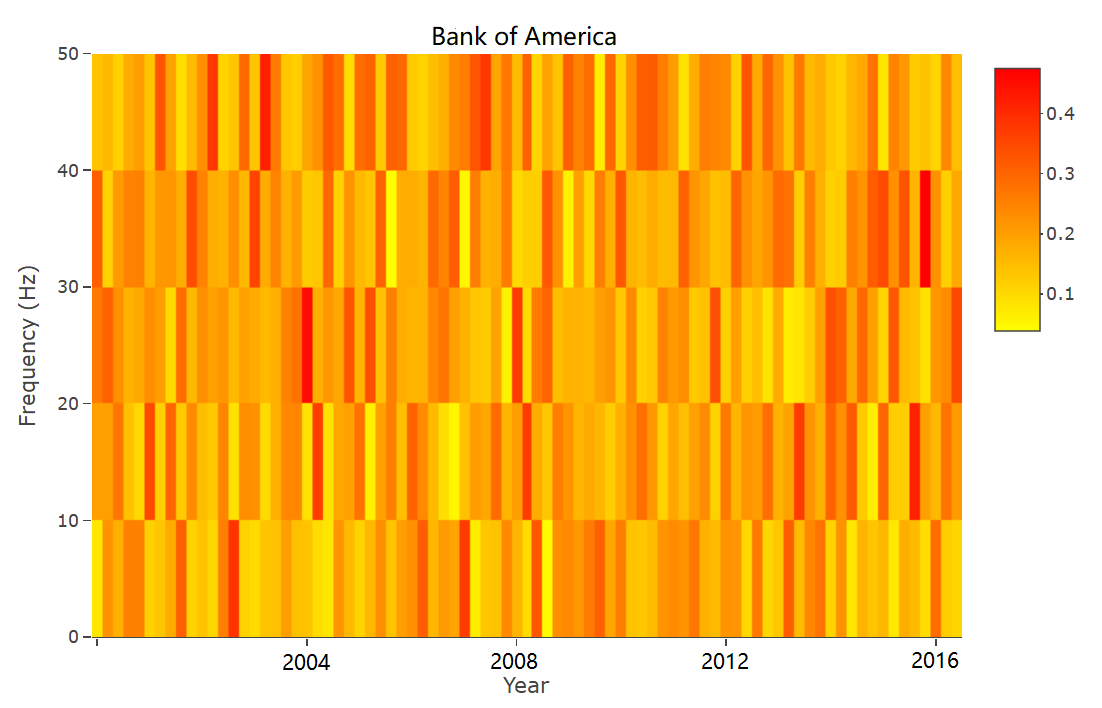}&
		\includegraphics[scale=0.37]{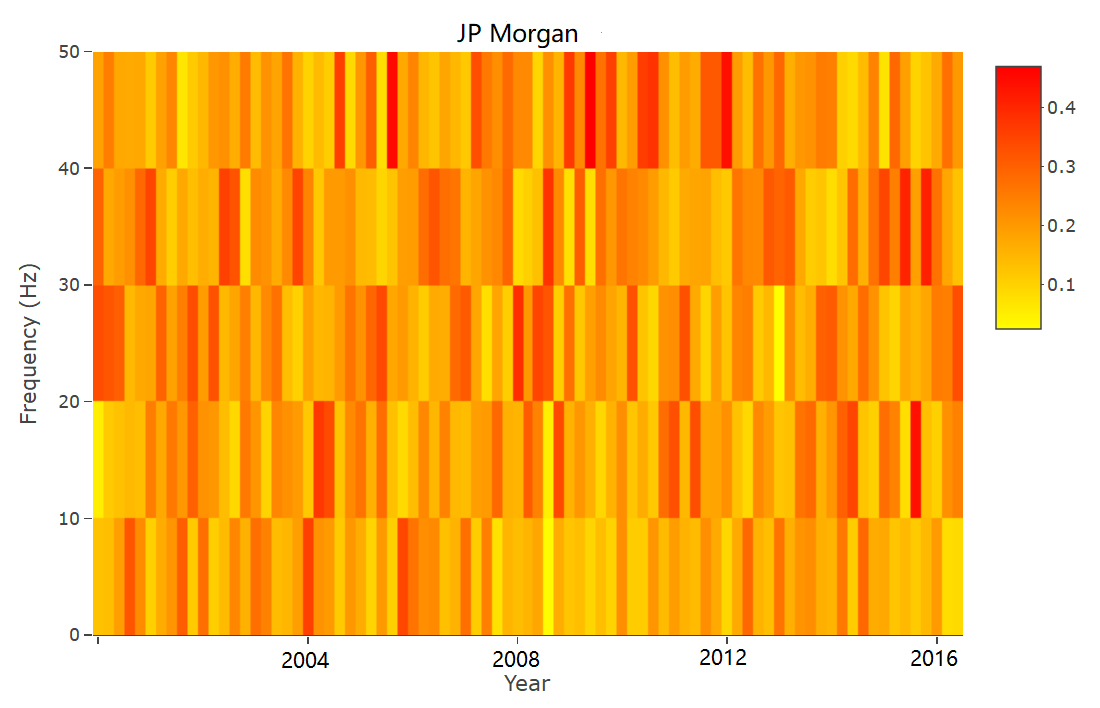}\\
		\includegraphics[scale=0.37]{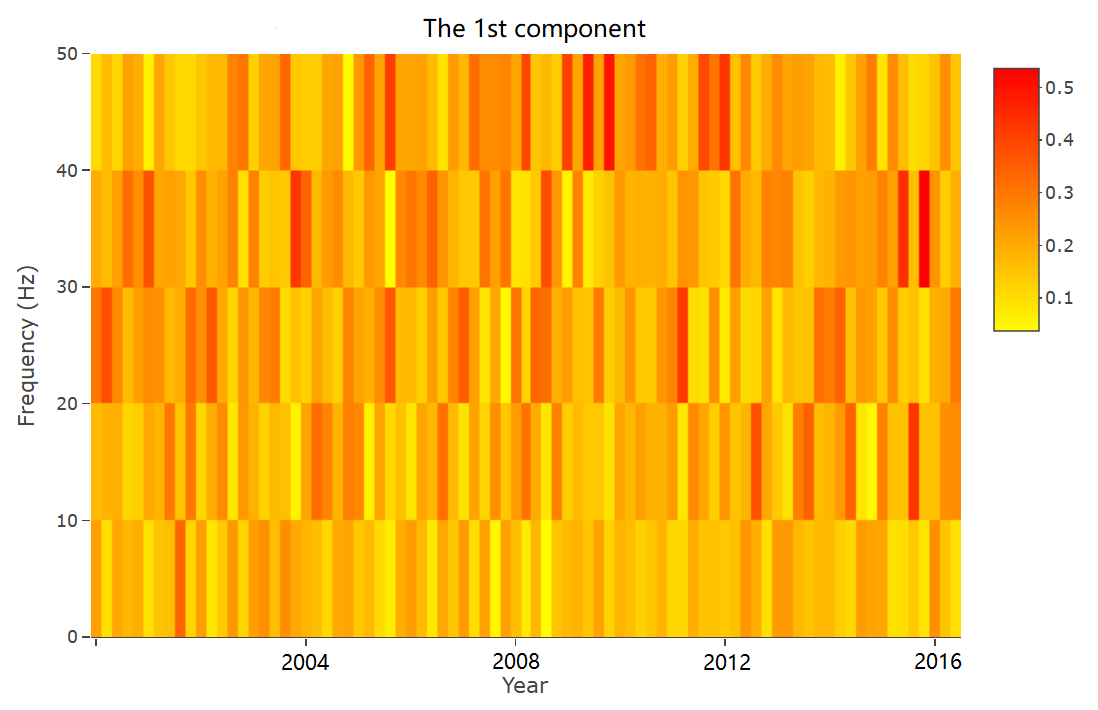}&
		\includegraphics[scale=0.37]{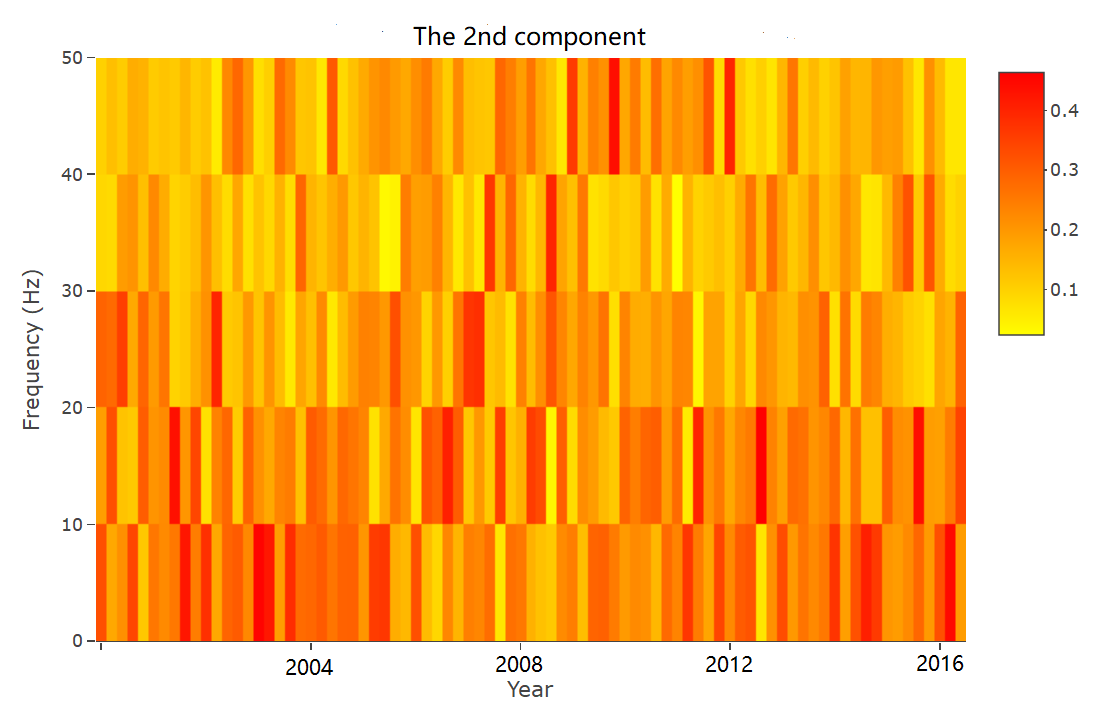}
	\end{tabular}
	\caption{ Time-frequency plots. Top: Bank of America and JP Morgan. Bottom: The first and second spectral PCs. The x-axis denotes the time; the y-axis denotes the frequency from 0 to 50 Hertz. }
\end{figure}

\begin{figure}[h!]
	\centering
	\begin{tabular}{c c}
		\includegraphics[scale=0.8]{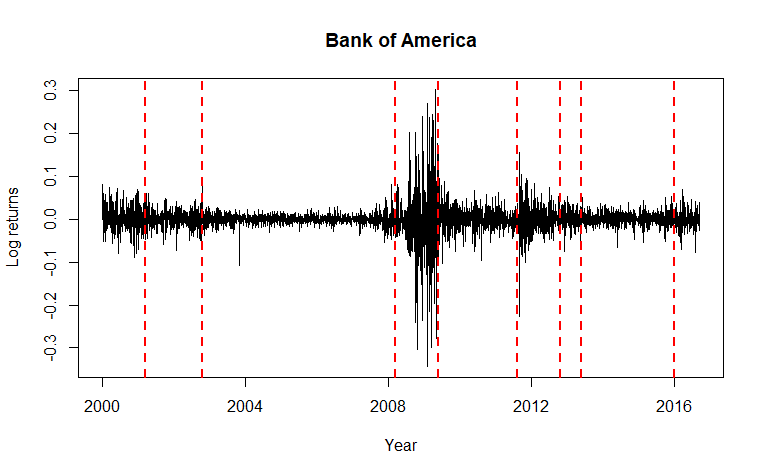}&
		\includegraphics[scale=0.8]{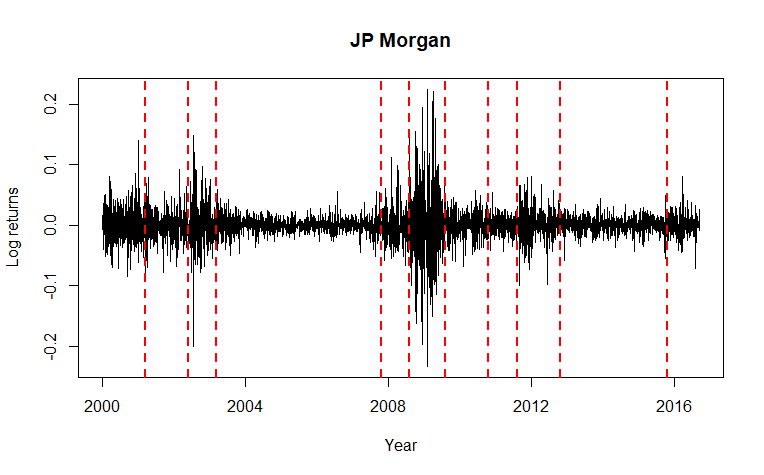}\\
		\includegraphics[scale=0.8]{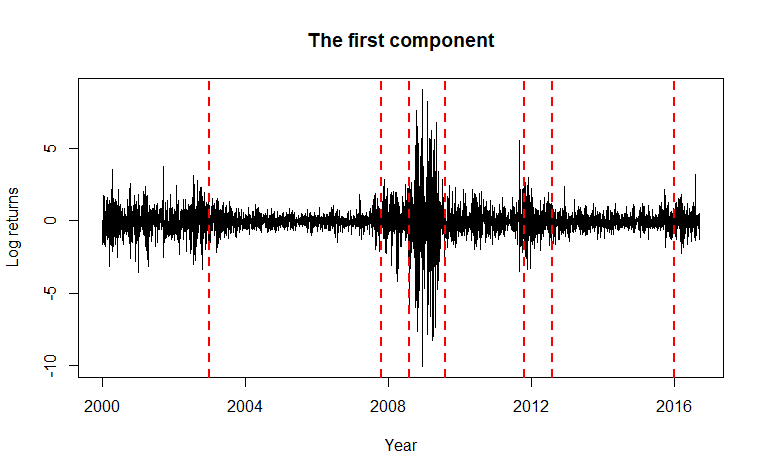}&
		\includegraphics[scale=0.8]{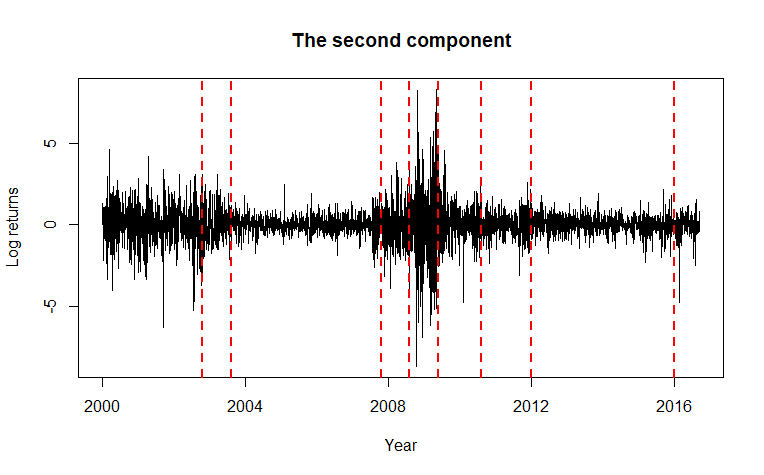}
	\end{tabular}
	\caption{Top: The time series of Bank of America and JP Morgan. Bottom: The first and second spectral PCs. The dotted lines denote the locations of estimated change points }
\end{figure}

\begin{figure}[h!]
	\centering
	\begin{tabular}{c c}
		\includegraphics[scale=0.8]{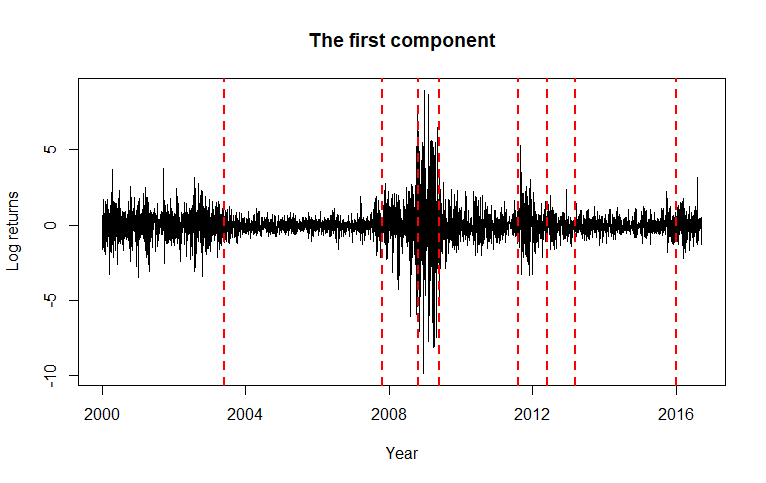}&
		\includegraphics[scale=0.8]{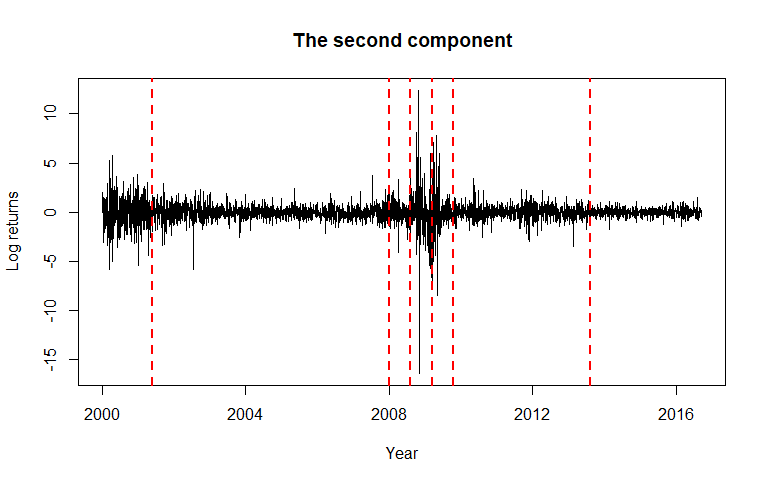}
		\\
		\includegraphics[scale=0.8]{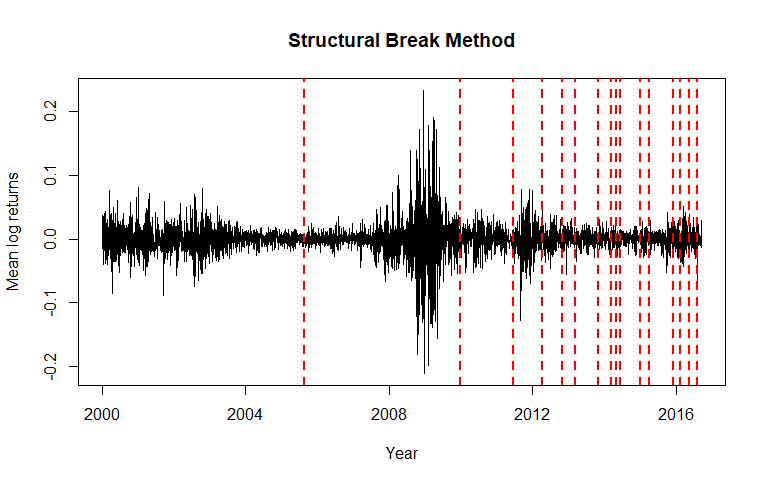}&
		\includegraphics[scale=0.8]{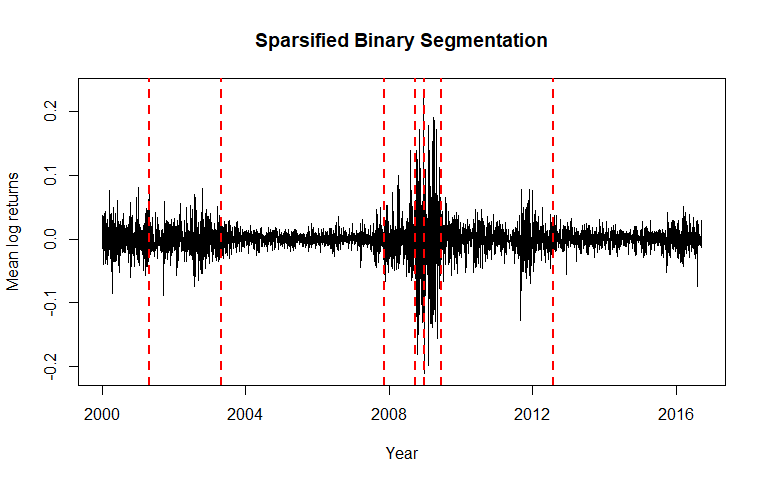}
	\end{tabular}
	\caption{Top: The estimated change point using the first or second component of Contemperaneous Method. Bottom: The estimated change point using Structural Break Method and Sparsified Binary Segmentation. The red dotted lines denote the location of estimated change points. }
\end{figure}

\FloatBarrier

\section{Conclusion}
In this paper, we proposed the Spec PC-PC method, a new change-point detection approach using multivariate time series such as EEG and stock data. The approach consists of two stages: using Spectral PCA to obtain a low-dimensional time series (stage 1) and then applying a binary segmentation algorithm to detect change points (stage 2). In the simulation studies, the spec PC-CP method outperforms competing methods in both detection rate and mean absolute difference. We compared our method with Contemporaneous Mixture method, Structural Break method and Sparsified Binary Segmentation under a variety settings.
These results confirm that Spec PC-CP can capture the time-lag effect in multivariate time series.

When we applied the method to the EEG seizure data, change points were identified at the beginning and around seizure onset. It suggests possibility to build early warning systems of seizure and gives more precise diagnosis, as changes are often not obvious to be observed through visual inspection. We also analyzed the stock data with daily log returns of several representative stocks. Change points are detected around events that have impact on the financial market. For example, the start of the Iraq War in 2003 and the financial crisis in 2008. It demonstrates that our method is efficient in analyzing financial data, which could lead to insights in understanding the fluctuation of financial market.

There are several potential future research directions. First, in simulations we look at the first three components. It is of interest to explore more components and understand how they capture the information of the original time series. Second, research has been done on forecasting time series. For example, Fryzlewicz (2005) modeled financial log-return series in the Locally Stationary Wavelet (LSW) framework. It will be interesting to combine Spectral PCA and forecasting methods to predict future events such as seizure and financial crisis. 

\newpage
\section{Appendix}
In the Structural Break simulation setting we have a piecewise stationary time series model with a single change point at $t=500$.
At each time point $t$, $X(t)$ is a $k\times 1$ vector. $\epsilon(t)\sim N(0, 0.01I_k)$. We have $p=10$ in this case.

\[ X(t) = \left\{
\begin{array}{ll}
0.9X({t-1})+\epsilon(t) & (0<t<=500)\\
-0.9X({t-1})+\epsilon(t) & (500<t<=1,000)
\end{array}
\right.
\]
\begin{table}[h!]
	\centering
	\caption{Summary of simulation results using Structural Break setting. The length of time series lengths is chosen as $T=1,000$. There is one change point at $t=500$.}
	\begin{tabular}{c c c}
		\hline
		& Detection rate  &  MAD \\
		\hline
		Spec PC CP   & 0.97 & 20.1 \\
		Contemporaneous Method   & 0.96 & 18.6 \\
		Structural Break   &1 & 1.14 \\
		Cho's Method   &0.93 & 2.05 \\
		\hline
	\end{tabular}
\end{table}

In the simulation setting in Cho and Fryzlewicz, (2015), we have a piecewise stationary time series model with a single change point at $t=500$ with $\alpha\sim Uniform(0.5,0.59)$ and $\beta\sim Uniform(-0.79,-0.5)$.
At each time point $t$, $X(t)$ is a $k\times 1$ vector. $\epsilon(t)\sim N(0, 4I_k)$. We have $k=100$ in this case.

\[ X(t) = \left\{
\begin{array}{ll}
\alpha X({t-1})+\epsilon(t) & (0<t<=500)\\
\beta X({t-1})+\epsilon(t) & (500<t<=1,000)
\end{array}
\right.
\]

\begin{table}[h!]
	\centering
	\caption{Summary of simulation results under Cho's setting. The length of time series lengths is chosen as $T=1,000$. There is one change point at $t=500$.}
	\begin{tabular}{c c c}
		\hline
		& Detection rate  &  MAD \\
		\hline
		Spec PC CP   & 0.99 & 21.2 \\
		Contemporaneous Method   & 0.98 & 22.3 \\
		Structural Break   &0.34 & 442.0 \\
		Cho's Method   &0.91 & 51.5 \\
		\hline
	\end{tabular}
\end{table}

\newpage
\nocite{*}
\bibliography{references}
\bibliographystyle{Chicago}

\end{document}